\documentclass{aa}               
\input psfig.sty
\def\simless{\mathbin{\lower 1pt\hbox
   {$\spose{\raise 5pt\hbox{$\char'074$}}\char'430$}}}
\def\simgreat{\mathbin{\lower 1pt\hbox
   {$\spose{\raise 5pt\hbox{$\char'076$}}\char'430$}}}
\def\simgreat{\gapp}
\def\simless{\lapp}
\def\lapp{\mathbin{\raise2pt \hbox{$<$} \hskip-9pt \lower4pt \hbox{$\sim$}}}
\def\gapp{\mathbin{\raise2pt \hbox{$>$} \hskip-9pt \lower4pt \hbox{$\sim$}}}
\begin{document}

   \title{Nonradial and nonpolytropic  astrophysical outflows\protect\\
         V. Acceleration and collimation of self-similar winds}
   \titlerunning{Nonradial and nonpolytropic  astrophysical outflows V.}

  \author{C. Sauty
           \inst{1}
   \and    E. Trussoni
           \inst{2}
   \and    K. Tsinganos
           \inst{3}
          }

   \offprints{C. Sauty \\ (christophe.sauty@obspm.fr)}

   \institute
         {Universit\'e de Paris 7 -- Observatoire de Paris, LUTH,
        unit\'e CNRS FRE 2462, F-92190 Meudon, France
    \and Istituto Nazionale di Astrofisica (INAF) - Osservatorio Astronomico 
       di Torino, Strada Osservatorio 20, I-10025 Pino Torinese (TO), Italy
    \and  Section of Astrophysics, Department of Physics, University of Athens, Panepistimiopolis 
         GR-157 84, Zografos, Greece
         }
   \date{Received 6 December 2001 / accepted 16 April 2002}

   \abstract{
An exact model for magnetized and rotating outflows, underpressured at their axis, is analysed 
by means of a nonlinear separation of the variables in the two-dimensional  
governing magnetohydrodynamic (MHD) equations for axisymmetric plasmas. 
The outflow starts subsonically and subAlfv\'enically from the central gravitating 
source and its surrounding accretion disk and after crossing the MHD critical 
points, high values of the Alfv\'en Mach number may be reached. 
Three broad types of solutions are found:  (a) collimated jet-type outflows 
from efficient magnetic rotators where the outflow is confined by the magnetic 
hoop stress; 
(b) collimated outflows from inefficient magnetic rotators where the outflow is 
cylindrically confined by thermal pressure gradients; and (c) radially expanding 
wind-type outflows analogous to the solar wind.  
In most of the cases examined cylindrically collimated (jet-type) outflows are 
naturally emerging with thermal and magnetic effects competing in the 
acceleration and the confinement of the jet.   
The interplay of all MHD volumetric forces in accelerating and confining the 
jet is displayed along all its length and for several parameters. The solutions may be used for a physical 
understanding of astrophysical outflows, such as those associated with young 
stellar objects, planetary nebulae, extragalactic jets, etc.  
}

\maketitle


\section{Introduction}

Rotating magnetized objects with hot coronae play a crucial role in
accelerating and collimating astrophysical outflows on the solar,
galactic and extragalactic scales. A magnetohydrodynamical
(MHD) approach is usually considered as the first and zeroth-order 
step for modeling such astrophysical outflows. 
However, the mathematical difficulties in the treatment
of the MHD equations are at such a high level that several
approximations are still unavoidable to obtain solutions useful
for an interpretation of the corresponding observed phenomena.  

Several efforts have been put forward recently in developing interesting 
and useful numerical simulations (e.g. Sakurai 1985, Ouyed \& Pudritz 1997, 
Bogovalov \& Tsinganos 1999,  Krasnopolsky et al. 1999,  Koide et al. 2000,  
Keppens \& Goedbloed 2000, Usmanov et al. 2000, Bogovalov \& Tsinganos 2001). 
Nevertheless, most of them usually apply only to one category of
object, such as the solar wind, jets from accretion disks, etc,  with a 
limited set of explored parameters and boundary conditions because the 
simulations are rather time-consuming.
Some of these simulations also do not necessarily produce complete 
solutions from the base of the outflow up to infinity. 
Others, describe only very transient features because they fail 
to converge towards a stationary or quasi-stationary state.  
And, there are still some doubts on how the boundaries of the numerical 
box may influence some of those simulations, especially in disk winds,  
a fact related to the problem of the correct crossing of the critical
surfaces that govern MHD flows (Vlahakis et al. 2000).  
Hence, analytical or semi-analytical investigations should be performed 
parallely to numerical simulations, in order to explore a more 
extended set of parameters, despite the necessary additional 
assumptions and simplifications.

In the framework of steadiness and axisymmetric geometry, the MHD
system reduces to the well known Bernoulli and transfield equations,
whose detailed analytical treatment is at a rather prohibitive level.  
General properties of these two equations can be studied analytically 
only in the asymptotic regime.  For example, in Heyvaerts \& Norman 
(1989) it has been shown that collimation into cylinders is a natural 
configuration for a high speed, superAlfv\'enic outflow, in the limit where 
the field is asymptotically force free but carries a net electric current.  
A complete solution connecting the base of the outflow to
its asymptotic zone, crossing all singularities, can be obtained
analytically only by neglecting the transfield equation and by 
integrating the ordinary differential Bernoulli equation along
a streamline. This allows the modeling of a 1-D flow around
specific regions, e.g., on the equatorial plane of a star (Weber \& 
Davis 1967, Belcher \& McGregor 1976), or, along the stellar rotational 
axis provided that the physical variables have been `a priori' averaged
across the flux tube (Lovelace et al. 1991).

An alternative approach largely followed in the last years for
a 2-D modeling of axisymmetric astrophysical outflows
is the well known {\it self-similar} technique. The basis of this
treatment is the assumption of a non-linear separation of the natural 
variables of the system of the MHD equations.  This effectively 
leads to a scaling law of one of the variables as function of one of 
the coordinates. The particular choice of the scaling variable depends 
on the specific astrophysical problem. 
General properties of self-similar MHD solutions are discussed in 
Vlahakis \& Tsinganos (1998, henceforth VT98; see also Tsinganos 
et al. 1996).

Since the well known paper of Blandford \& Payne (1982), solutions
self-similar in the radial direction have been investigated to
analyse the structure of winds from accretion disks (Ostriker
1997, Lery et al. 1999, Casse \& Ferreira 2000b, Vlahakis et al.
2000, and references therein, Aburihan et al. 2001). In most of these models, not valid
along the polar axis, the driving  and collimation processes
derive from a combination of the magnetic and centrifugal forces.
However, the presence of a hot disk corona can also help to drive 
the outflow very efficiently, a factor which may change drastically 
the initial launching conditions (Casse \& Ferreira 2000a).

In a series of studies we have analysed a class of MHD solutions
that are self-similar in the meridional direction (Sauty \& Tsinganos 
1994, henceforth ST94, Trussoni et al. 1997, henceforth TTS97, Sauty 
et al. 1999, henceforth STT99, and references therein). 
Such a treatment is complementary with respect to the corresponding 
radially self-similar one because it allows to study the physical 
properties of outflows close to their rotational axis. 
As in this region the contribution to the acceleration by the 
magnetocentrifugal forces is small, the effect of a thermal driving 
force is very important.
Within this model, we either prescribe the meridional structure 
of the streamlines, or we assume a relationship 
between the radial and longitudinal components of the gas pressure 
gradient. 
The main properties of the first class of solutions which are 
asymptotically collimated are outlined in TTS97 where 
the essential role of rotation in getting cylindrical collimation is 
shown.

If the two components of the pressure gradient are related, the
meridional structure of the streamlines is deduced from the solution 
of the MHD equations.  Rotating magnetized outflows with a spherically
symmetric structure of the gas pressure may be asymptotically
superAlfv\'enic with radial or collimated fieldlines, depending on
the efficiency of the magnetic rotator (ST94). This allows to deduce a
criterion for selecting spherically expanding winds from cylindrically 
collimated jets. In STT99 we extended the results of ST94 by
performing an asymptotic analysis of the meridional self-similar
equations assuming a non spherically symmetric structure for the
pressure. It was pointed out that a superAlfv\'enic outflow may
encounter different asymptotic conditions: it can be thermally or
magnetically confined, and thermally or centrifugally supported.
Thus in the frame of this model, the conclusion stated in Heyvaerts 
\& Norman (1989) has been extended to cases which are not
asymptotically force-free, but also include pressure gradients
and centrifugal terms. Current carrying flows are cylindrically
collimated (around the polar axis, not necessarily everywhere in
the flow) for under-pressured flows. While overpressured flows
attain asymptotically both radial and cylindrical shapes.

We complete here this study  by seeking complete solutions 
that connect the basis of the flow with its superAlfv\'enic
regime. In particular we investigate if, and under which
conditions, the basal region can be matched to the asymptotic
solutions outlined in STT99. As this analysis implies a careful
topological study of the MHD self-similar equations, with a proper
treatment of the critical points, in this article we only discuss the
structure of {\it underpressured} outflows 
($\kappa>0$, see Eq. \ref{pressure}) that may be
asymptotically confined by an external pressure or a toroidal
magnetic field. The study of solutions for pressure supported jets
($\kappa<0$) 
is postponed to a forthcoming article.

In Sec. 2 we outline the main properties of the 
meridionally self-similar MHD equations while their 
asymptotic properties analysed in STT9 are summarized in Sec. 3. 
In Secs. 4 and 5 we discuss the main features of the numerical solutions
and their physical validity, while in Sec. 6 we analyse the dynamical
properties of the outflow. The general astrophysical 
implications of our investigation are discussed in Sec. 7.

\section{Governing equations for meridional self-similar outflows}
We summarize here the main assumptions in a meridionally ($\theta-$)
self-similar treatment of the MHD equations. 
More details may be found in STT94 and STT99, while a brief   
discussion of the various classes of self-similar MHD solutions 
is presented in Appendix A.

\subsection{General properties of axisymmetric steady flows \label{sec21}}

The basic equations governing plasma outflows in the framework 
of an ideal MHD treatment are the  momentum, mass and magnetic flux
conservation, the frozen-in law for infinite conductivity 
and the first law of thermodynamics. 
In steady and axisymmetric conditions (Tsinganos 1982, 
Heyvaerts \& Norman 1989), the MHD equations allow the 
following definition of the magnetic field $\vec{B}$ and velocity
$\vec{V}$ through the poloidal magnetic flux function $A(r, \theta)$ and
the poloidal stream function $\Psi(r,\theta)$, in spherical coordinates 
($r, \theta, \varphi$)
\begin{equation}\label{B}
\vec{B}= {\vec \nabla A \over r \, \sin \theta } \times 
\hat \varphi\, - \, \left [ {L \Psi_A \over r\, \sin \theta} {1 -
r^2 \, \sin^2 \theta \, \Omega/L \over 1 - M^2} \right ] \hat \varphi 
\,, 
\end{equation}
\begin{equation}\label{V}
\vec{V} = {\Psi_A \over 4 \pi \rho}
\vec{B}_p \, + \, \left [ {L  \over r \, \sin \theta} \, {r^2 \,
\sin^2 \theta \, \Omega/L -M^2 \over 1 - M^2} \right ] \hat \varphi 
 \,,
\end{equation}
\noindent
where $\Psi_A = {\mathrm d} \Psi / {\mathrm d} A$. $L(A)$ is the total angular
 momentum carried by the flow along a fieldline $A$ = const, $\Omega(A)$ is
the angular velocity of this fieldline at the base of the flow, and
\begin{equation}\label{M}
M^2 = 4 \pi \rho {{v^2_p} \over {B^2_p}} = {{\Psi^2_A} \over {4 \pi
\rho}}
\,, 
\end{equation}
\noindent is the square of the poloidal Alfv\'en number. From Eqs.
(\ref{B}) and  (\ref{V}) we see that for transAlfv\'enic outflows 
 $L(A)$ and $\Omega(A)$ are not independent integrals: for  $M=1$,
 we must have
$L(A)/\Omega(A)= [r^2 \, {\sin}^2 \theta]_{M=1}$.

Taking into account Eqs. (\ref{B}) and (\ref{V}), the original system of 
MHD equations reduces to two coupled partial differential equations for 
the density $\rho$ and the flux function $A$.

\subsection{Self-similarity: scaling laws of the present model \label{sec22}}

There are two key assumptions in a meridionally self-similar 
treatment of the MHD equations.

The {\it first} is that the Alfv\'en number is solely a 
function of the radial distance, $M \equiv M(r)$.  
It is then convenient to normalize all quantities on the Alfv\'en 
surface along the rotation axis, $r=r_*$.    
Then denote the dimensionless radial distance by $R=r/r_*$, 
while $B_*$, $V_*$ and $\rho_*$ are the poloidal magnetic field, 
velocity and density along the polar axis at the radius $r_*$, 
with $V^2_*= B^2_* / 4 \pi \rho_*$. 
Also, define the dimensionless magnetic flux function 
$\alpha(R, \theta) = 2 \, A(r, \theta) / r^2_* B_*$.  

The {\it second} is  the assumption of a dipolar dependence of the 
magnetic flux function $A$:
\begin{equation}\label{alpha}
A(r, \theta) = A(\alpha),\;\;\;\;\;\;\;\;\;\alpha = {{R^2} \over
          G^2(R) } {\rm sin}^2 \theta
\,,
\end{equation}
\noindent
where $G^2 (R)$ is the cross sectional area of a flux tube 
perpendicularly to the symmetry axis, in units of the corresponding 
area at the Alfv\'en distance. 
Then, for a smooth crossing of the Alfv\'enic surface the regularity 
condition for the toroidal components of the magnetic and velocity 
fields becomes $L(A)/\Omega(A) = r_*^2 \alpha$. 

The next step of the method is to choose the two free functions 
of $\alpha$ ($\Psi_{A}$ and $\Omega$) such that the 
variables 
($R, \theta$) separate.

\noindent
{\underbar{\it Ratio of magnetic to mass flux.}} 
We assume a linear dependence  of
$\Psi^2(\alpha)$, which also fixes the density profile, Eq. (\ref{M}):
\begin{equation}\label{density}
\label{rho}
\Psi_A^2=4\pi\rho_*(1 + \delta \alpha)
\,,\qquad
\rho(R,\alpha) = {{\rho_*} \over {M^2(R)}} (1 + \delta \alpha)
\,.
\end{equation}
The  parameter $\delta$ governs the non spherically symmetric
distribution of the density. This means
to assume a linear increase (or decrease) of the density when receding from
the rotational axis.

\noindent {\underbar{\it Total angular momentum and corotational
speed}}. For the poloidal current within a surface labeled by
$A(\alpha)=$ const [$\propto L(\alpha) \Psi_A(\alpha)$] we choose a
linear dependence on $\alpha$ through
 the parameter  $\lambda$.  Then from the regularity conditions in Eqs. (\ref{B})
and (\ref{V}), and the form of $\Psi_A(\alpha)$ we get
\begin{equation}\label{LOM}
L(\alpha) = \lambda r_* V_* {\alpha \over {\sqrt{1 + \delta \alpha}}},
\;\;\;\;\;\;\;\;\Omega(\alpha) = \lambda {{V_*} \over {r_*}} {1 \over
{\sqrt{1 + \delta \alpha}}}
\,.
\end{equation}
\noindent
Note that $\lambda$ is related to the rotation of the  poloidal
streamlines at $R=1$.

\noindent {\underbar{\it Pressure of the gas}}. Consistently with
the scaling of the density, the pressure of the gas is assumed
to vary linearly with $\alpha$ through the constant $\kappa$:
\begin{equation}\label{pressure}
P(R,\alpha) = {1 \over 2} \rho_* V^2_* \Pi(R)[1+ \kappa \alpha]\,.
\end{equation} 
For $\kappa> 0$ ($\kappa < 0$) the gas pressure  increases  
(decreases) by moving away from the polar axis. 
We remark that this assumption substitutes the usual polytropic 
relationship. Eq. (\ref{pressure}) is effectively equivalent with 
a relationship between the heating rate $q$ in the gas, $P$ and $A$. 
A special case of such a functional relationship among $q$, $\rho$, $P$ and
 $A$ yields the familiar polytropic law $P \propto \rho^{\gamma}$ (ST94).

In this paper we confine our attention to underpressured jets, i.e., when 
$\kappa> 0$.

\noindent
{\underbar{\it Gravity}}. Consequent to the scaling of the density, the 
gravitational force per unit volume is
\begin{equation}
{\vec f}_g = - \rho {{\cal G M}  \over r^2} \, \hat r
= -{1 \over 2} \rho_* {V^2_* \over r_*}
{\nu^2 \over M^2 R^2 }(1+\delta \alpha )
\,,
\label{grv1}
\end{equation}
where ${\cal M}$ is the central gravitating mass and 
we have introduced a new parameter $\nu$ which is the ratio of 
the  escape 
velocity to the flow velocity on the polar axis at $R=1$
\begin{equation}
\nu^2 ={{\cal G M}   \over r_* V^2_*}
\,,
\label{grv2}
\end{equation}

\noindent
{\underbar{\it The expansion factor}}. For homogeneity with the
notation in  ST94 and STT99  
we introduce the  function $F(R)$, which is the logarithmic 
derivative (with a minus
sign) of the well known expansion factor used in solar wind
theory (Kopp \& Holzer 1976):
\begin{equation}\label{F}
F(R) = 2 \, \left [1 -
{{\mathrm d} \, \ln G(R) \over {\mathrm d} \, \ln R} \right ]
\,.
\end{equation}
\noindent We remind that the value of $F$ defines the shape of the
poloidal streamlines (that are parallel to the poloidal magnetic
fieldlines). For $F(R) =  0$ the streamlines are radial (as e.g.
in Tsinganos \& Trussoni 1991), for $F(R) > 0$ they are deflected
to the polar axis (and $F=2$ means cylindrical collimation, see 
ST94 and TTS97), while for $F(R) < 0$  they flare to the
equatorial plane.

 From the previous definitions we deduce the components of $\vec{V}$ 
and $\vec{B}$ as functions of $R$, $\theta$, $G(R)$, $F(R)$ and $M(R)$. 
Then, the momentum  conservation law provides three ordinary 
differential equations which together with Eq. (\ref{F}) can be solved 
for the four variables $M^2(R)$, $F(R)$, $\Pi(R)$ and $G(R)$ 
(see Appendix B). 
For $\kappa \geq 0$ these equations have two singularities: on the
Alfv\'en surface and at the position where the radial component of
the flow velocity is equal to the radial component of the slow
magnetosonic velocity (for a general discussion of the 
singularities of the self-similar MHD equations see Tsinganos et al.
1996). The details of the regularity conditions that must be
fulfilled to cross these critical surfaces are widely discussed in
ST94. They are a crucial element to ensure well posed boundary
conditions (Sauty et al. 2001).

\subsection{Energetics}

The conserved total energy flux density per unit of mass flux density 
$E(\alpha)$ is equal to the sum of the poloidal kinetic, rotational and 
gravitational energies, together with the enthalpy and heating along a specific 
streamline. In the framework of the present meridionally self-similar 
model, $E(\alpha )$ can be expressed as (ST94):
\begin{equation}\label{En2}
E(\alpha)= {1 \over 2} V^2_* {{{\cal E} + \alpha \Delta {\cal E}}
\over  {1 + \delta \alpha}}
\,.
\end{equation}
The two constants ${\cal E}$ and $\Delta {\cal E}$ appearing in 
Eq. (\ref{En2}) represent the specific
energy along the polar streamline and the variation of the specific energy
across the streamlines, respectively (see STT99).
Due to the assumed linear dependence of the pressure
with $\alpha $, Eq. (\ref{pressure}), it turns out that the following quantity
\begin{eqnarray}
\epsilon  = \Delta {\cal E} - \kappa {\cal E}
\qquad\qquad\qquad\qquad\qquad\qquad\nonumber\\
={M^4\over (GR)^2}\left[ {F^2\over 4} - 1 \right]
- \kappa {M^4\over G^4}
- {(\delta\,- \kappa) \nu^2 \over R}
\nonumber\\
\label{EnEps}
+ {\lambda^2 \over G^2} \left({M^2-G^2 \over 1-M^2}\right)^2
+ 2\lambda^2{1-G^2 \over 1-M^2}\label{epsilon}
\end{eqnarray}
is a constant on \underbar{all} streamlines (ST94). Physically, $\epsilon$ is
related to the \underbar{variation} across the fieldlines of the specific energy 
which is left available to collimate the outflow once the thermal content
converted into kinetic energy and into balancing gravity has
been subtracted (STT99). For $\epsilon >0$ collimation
is mainly provided by magnetic terms, while for $\epsilon <0$ the
outflow is confined mainly by thermal pressure. Accordingly, in STT99
we defined flows with positive or negative $\epsilon$ as {\it
Efficient} or {\it Inefficient} Magnetic Rotators, respectively
({\bf EMR} or {\bf IMR}).

Note that $\epsilon$ is not a mere generalization of the Bernoulli
constant but rather a variation of the total energy per unit mass 
from one magnetic fieldline to the next. 
Therefore it also contains information on the transfield force 
balance equation and the energies that control the shape of the 
flow tube rather than only information on the various driving mechanisms.
We also note that an analogous constant does not exist in 
the case of meridionally self-similar solutions with prescribed 
streamlines discussed in TTS97, where Eq. (\ref{pressure}) 
does not hold. 

\section{Asymptotic behaviour of the solutions\label{sec3}}

For $R \gg 1$ the asymptotic parameters of collimated outflows 
($F_{\infty} =2$, $G_{\infty}$ and $M_{\infty}$ bounded) depend 
on the value of $\epsilon$ and the force balance across the poloidal 
streamlines,  $f_{\nabla P} + f_B + f_C=0$, with  
$f_{\nabla P}$, $f_B$ and $f_C$ the pressure gradient, 
magnetic stress and centrifugal force, respectively.
Then,  we may calculate $M_{\infty}$ and $G_{\infty}$ as
functions of the parameters $\epsilon/2 \lambda^2$, $\kappa/2
\lambda^2$ and $\Pi_{\infty}$.

The asymptotic properties of these self-similar winds have 
been discussed in detail in STT99, and here we summarize 
their main features for the case of underpressured outflows 
($\kappa > 0$).

\begin{itemize}
\item{} Two main confining regimes exist. In one the outflow 
is collimated by the pinching  of the toroidal magnetic 
field ($\epsilon > 0$, {\it magnetic} regime: EMR) and in the 
other by the thermal pressure  ($\epsilon < 0$,
{\it thermal} regime: IMR).

\item{} For $\kappa \rightarrow 0$ we have collimation only 
for $\epsilon > 0$:
there is no pressure gradient across the streamlines and the flow can be
confined only by the magnetic stress ($f_B+f_C=0$)

\item{} The flow is supported by the centrifugal force and
collimated either by the thermal pressure ($\epsilon/2 \lambda^2 \ll 0$, $f_C +
f_{\nabla P}=0$), or by the magnetic pinch ($\epsilon/ 2 \lambda^2 \sim 0$, $f_C +
f_B=0$).

\item{} The collimated streamlines always show oscillations. This behaviour is
consistent  with  the results  found in more general, non self-similar
treatments (VT98).
\end{itemize}
In conclusion and within the present model, from the asymptotic 
analysis it turns out that underpressured and meridionally self-similar 
outflows should be in principle always collimated.  However, 
for very small values of $\kappa$ and asymptotically vanishing pressure
$\Pi_\infty=0$, an asymptotically radial configuration of the fieldlines is not
excluded.

\section{Numerical results\label{sec4}}

\subsection{Numerical technique and parameters}\label{subs41}

Using routines of the NAG scientific package 
suitable for the treatment of stiff systems and the Runge-Kutta
algorithm, Eqs. (\ref{F}), (\ref{Eq1}) - (\ref{Eq7}) are
integrated upstream and downstream of the vicinity of the
Alfv\'en point ($R_{in} = 1 \pm {\rm d} R$) with $M_{in}=1 \pm p
\, {\rm d}R$ and $G_{in} = 1 \pm (2 - F_{in}) {\rm d}R$ ($F_{in}
\approx F_*$). The slope $p$ of $M$ at $R=1$ is given in Eq.
(\ref{Eq8}). We first integrate upstream tuning the value of
$F_{in}$ until we select the critical solution that smoothly
crosses the singularity and reaches the base of the wind $R_o$,
where $M \rightarrow 0$. With this value of $F_{in}$ we then
integrate downstream to the asymptotic region ($R_{\infty}$ usually between
$10^4$ and $10^6$) and we get the complete solution. The value of the pressure
$\Pi_{in}$ $(\approx \Pi_*)$ is chosen such that $\Pi (R)$ is positive
everywhere (see  however Sec. 5 for a possible release of this
constraint).

We have seen in Sec. \ref{sec3} that the asymptotic properties of collimated
outflows are ruled by the three parameters $\epsilon/ 2
\lambda^2$, $\kappa/ 2 \lambda^2$ and $\Pi_{\infty}$. As it is
evident from Eqs. (\ref{Eq1}) - (\ref{Eq3}) and (\ref{epsilon}), the
numerical solutions require five parameters: $\epsilon$, $\kappa$,
$\lambda$, $\delta$ and $\nu$.
 However only four of them
are independent due to the constraint imposed by the integral
$\epsilon$, Eq. (\ref{EnEps}), which has the following expression at $R=1$:
\begin{equation}\label{epsilon2}
\epsilon = (\kappa - \delta) \nu^2  +\lambda^2 (\tau^2 + 1) - (1 -
\kappa) + F^2_*/4 \,.
\end{equation}
where $\tau [= (2 - F_*)/p]$ is given by Eq. (\ref{Eq8}).

\begin{figure*}
\centerline{
\psfig{figure=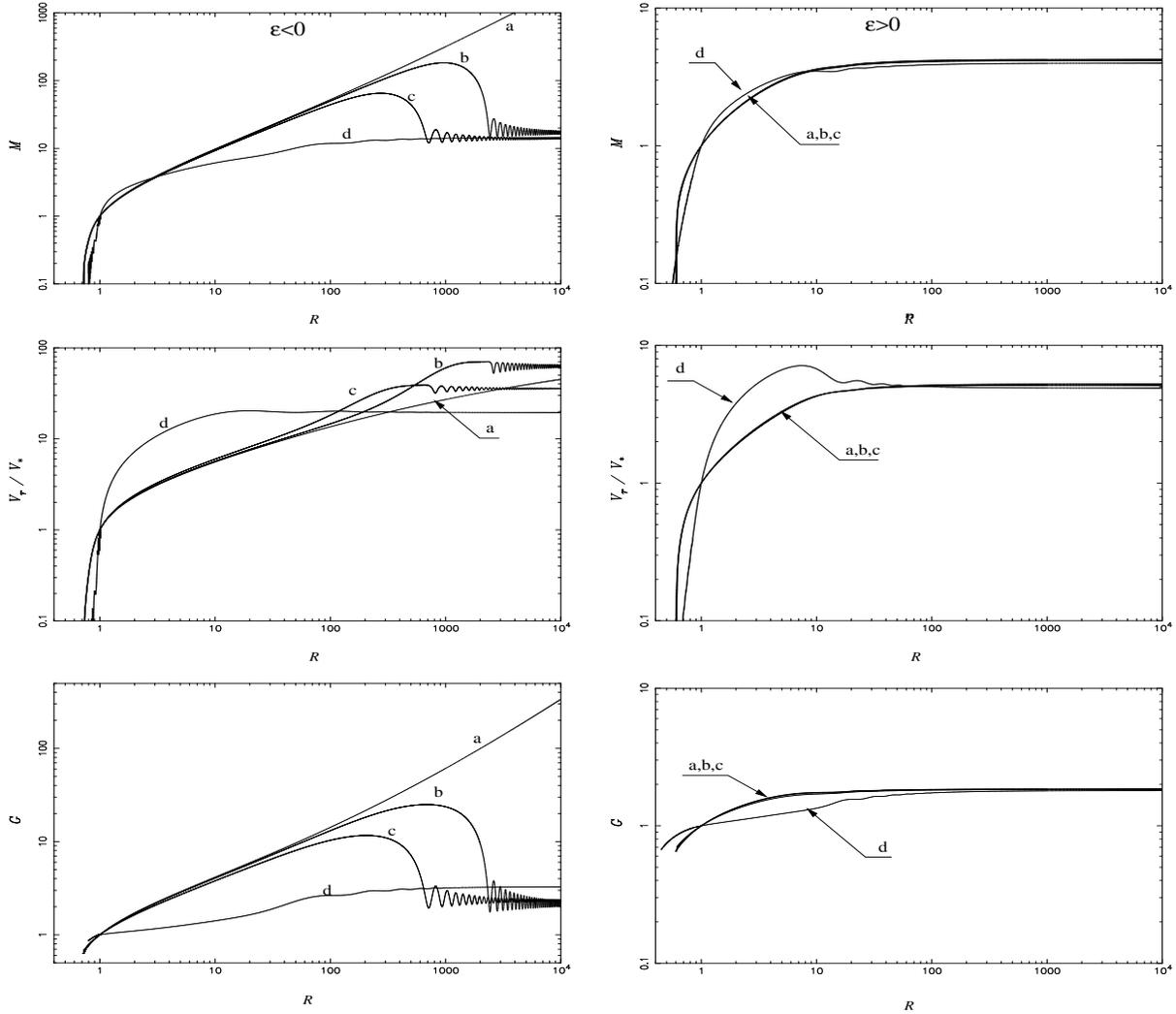,height=14.0truecm,width=16.0truecm,angle=0}}
\caption{Plots of $M$ (upper panel), $V_r$ (in units of $V_*$, middle panel) and $G$
(lower panel) vs $R$ along the polar axis for $\epsilon/2
\lambda^2 = -0.1$ (left panel)  and $0.25$ (right panel). Labels
refer to different values of $\kappa = 0$ (a), 0.001 (b),
0.003 (c) and 0.009 (d). In all these solutions we have
selected the solution corresponding to $\Pi_* = \Pi_{*,\rm min}$.
\label{f1}}
\end{figure*}

\begin{figure*}
\centerline{
\psfig{figure=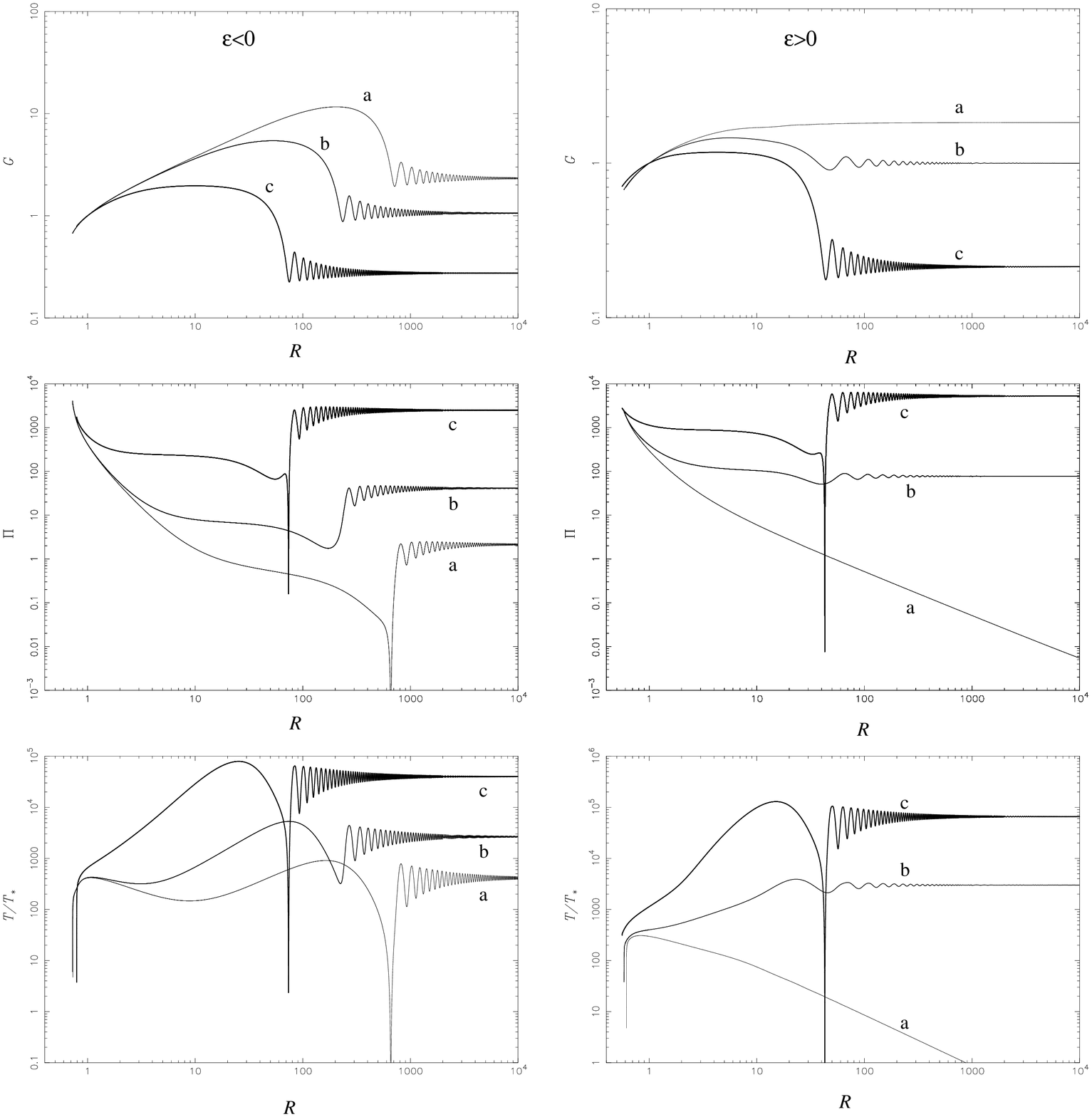,height=14.0truecm,width=16.0truecm,angle=0}}
\caption{Plots of $G$ (upper panel), $\Pi$ (middle panel) and $T$
(lower panel) vs $R$ along the polar axis for $\kappa = 0.003$ and
 $\epsilon/2 \lambda^2 = -0.1$ (left column) and
$\epsilon/2 \lambda^2 = 0.25$ (right column).
Labels refer to different values of $\Pi_*$. 
For  $\epsilon/2 \lambda^2 = -0.1$, $\Pi_*
\approx 413$ (a, $\Pi_{*,\rm min}$), 420 (b), 653 (c, $\Pi_{*,max}$).
For $\epsilon/2 \lambda^2 = 0.25$, $\Pi_* \approx 297$ (a,
$\Pi_{*,\rm min}$), 400 (b), 1135 (c, $\Pi_{*,max}$). 
\label{f2}}
\end{figure*}

We remind that the parameters $\lambda$, $\delta$ and $\kappa$ are
related to the strength of the rotational velocity, and to the
structure of the density and of the pressure in the $\theta$
direction, respectively. For $\delta$ and $\kappa$ positive the
density and pressure increase when receding from the polar axis.
From the expression giving the base of the outflow $R_o$ 
as deduced from the  definition of $\epsilon$ with $M_0=0$  
and $G_0 \ll 1$, Eq. (\ref{epsilon2}):

\begin{equation}\label{Ro}
R_o = {{(\delta - \kappa) \nu^2/\lambda^2} \over { 2  - G^2_o -
\epsilon/\lambda^2}} \,,
\end{equation}

\noindent
it is evident that the quantity $(\delta-\kappa) \nu^2$ rules the flow
dynamics in the subAlfv\'enic region.  Large
values of it lead to steep initial acceleration, with the base of the
flow close to the Alfv\'en surface. 
If $\delta=\kappa  $, $R_0 =0$ and there is no acceleration close to 
the base for a given speed at the Alfv\'en transition.  
We also see that we must have $\kappa < \delta$ in order that $R_0 > 0$ while 
$\delta >0$ ($\delta <0$) increases (decreases) the initial acceleration. 
For $\kappa > \delta$ the initial thermal driving of the outflow is not
possible at the base (an analogous case was found in TTS97, where
the quantity $\delta \nu^2$ had to be larger than a minimum
threshold value to have mass ejection).

In the following two subsections we
have fixed $\lambda=3$ and $\delta= 0.01$ [with the value of $\nu$
deduced from Eq. \ref{epsilon2}] and analysed the trend of the solutions for
different values of $\epsilon/2 \lambda^2$, $\kappa / 2 \lambda^2$
and $\Pi_*$. A summary of all these properties is given in 
subsection (\ref{subs44}) for $\lambda=\nu=1$.

\subsection{Behaviour of the solutions with $\kappa$}\label{subs42}

In Figs. \ref{f1} we have plotted the Alfv\'en number $M(R)$, the radial 
velocity $V_r(R)$ along the polar axis and the jet radius $G(R)$ for $\epsilon/ 2
\lambda^2=0.25$ and $-0.1$, and four different values 
of $\kappa =$ 0, 0.001, 0.003, 0.009. 
The chosen value of $\Pi_*$ is the minimum allowed
($\Pi_{*,\rm min}$), i.e. such that $\Pi$ is either vanishing asymptotically, 
or it is zero in the deepest minimum of the corresponding
oscillations of the streamlines. With these assumptions we deduce
the values of $\nu \approx 20 \div 100$. The main parameters 
of  the solutions are listed in Table \ref{tab:tab1}.

In the subAlfv\'enic region ($R \lapp 1$), independently of the
value of $\epsilon$, the increase of $\kappa$ rises the value of
$F_*$ (i.e. the rate of expansion of the streamlines at $R=1$ is
lower) and of the pressure: the outflow is `hotter' and `narrower'
in the transAlfv\'enic region. For $\kappa = 0 \rightarrow 0.009$ 
$\Pi_{*,\rm min} \approx 200 \div 2000$ (see Table
\ref{tab:tab1}). These effects however are evident only for
$\epsilon < 0$, while for positive $\epsilon$ the solutions are
practically unaffected by the value of $\kappa$ (unless $\kappa
\rightarrow \delta$, in which case oscillations of small amplitude
occur and $R_o$ decreases). 
The location of the base of the flow $R_o$ is controlled by $(\delta -
\kappa)\nu^2$. However in our plots the base of the flow does not
change drastically except when $\kappa \rightarrow \delta $
(see Table \ref{tab:tab1}). These trends are
basically unaffected by the value of $\epsilon/2 \lambda^2$.

In the superAlfv\'enic region ($R \gapp 1$), when $\epsilon > 0$ 
the results are basically insensitive to an increase 
of the values of $\kappa $.  The flow is always collimated as in ST94, while   
the radius of the jet is {\it smoothly} increasing.  
On the other hand, when $\epsilon<0$, the effect of $\kappa \neq 0$ 
is much more drastic.  
The outflow rapidly expands with $R$, reaches a maximum width 
and then it sharply recollimates to a rather narrow jet with large 
amplitude oscillations. For very small values of $\kappa$ ($\ll \delta$)
this bump increases its amplitude while its position moves downwind. 
In the limiting case of $\kappa \rightarrow 0$, the bump is shifted to 
$R= \infty$ and the flow has radial streamlines. 
For  $\kappa \rightarrow \delta$ the bump and the oscillations have 
shifted upstream and smoothed, and the solution becomes similar 
to the one  with $\epsilon >0$. 
Note also that at the position of the bump the Alfv\'en
number has a maximum (and the density is minimum), while the
velocity is monotonically increasing and attains asymptotic values
larger than in solutions with $\epsilon >0$.

\begin{figure*}
\centerline{
\psfig{figure=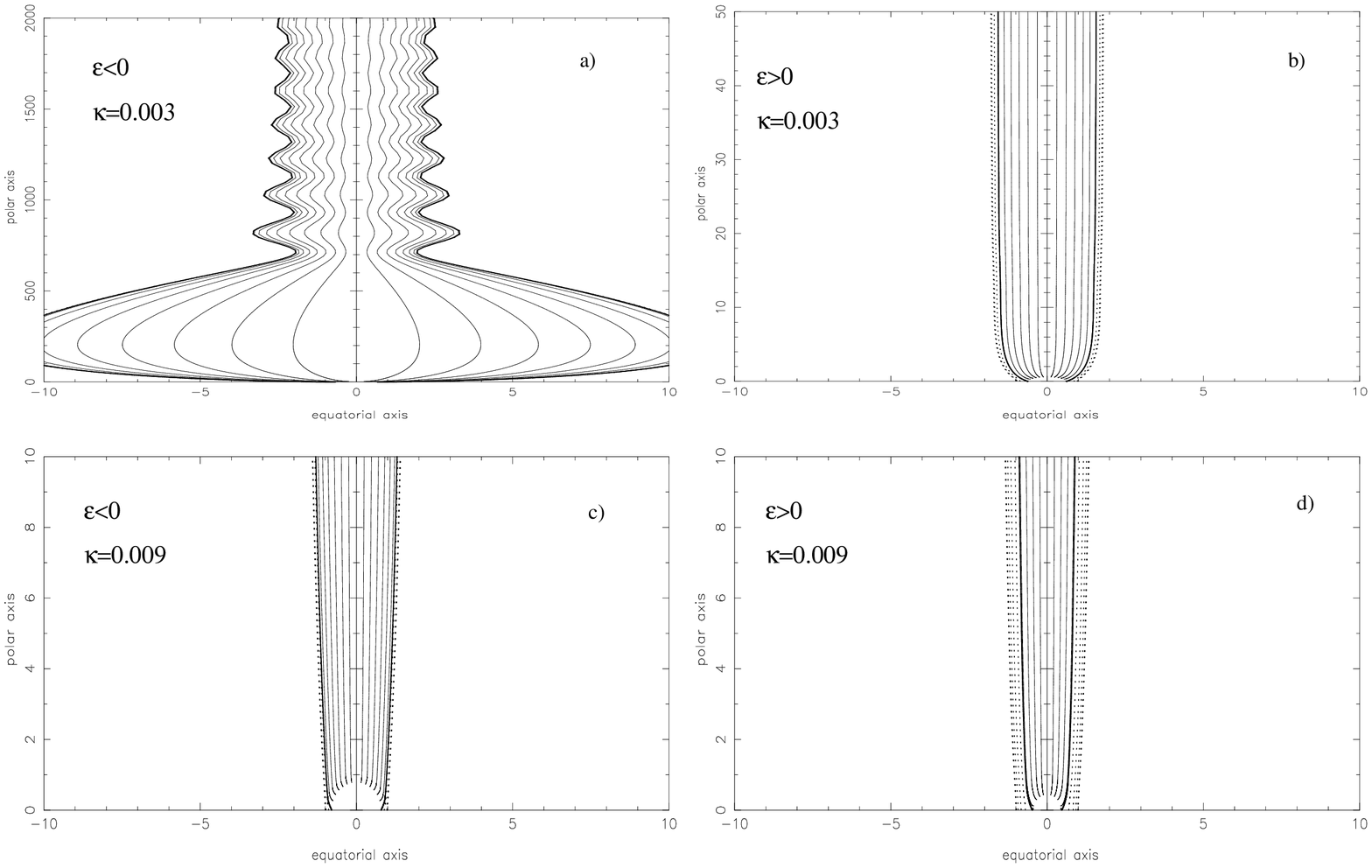,height=12.0truecm,width=16.0truecm,angle=0}}
\caption{Poloidal streamlines of some of the solutions discussed in Sec. \ref{sec4} 
and in Fig. (\ref{f1}). The last solid line corresponds to the last  streamline which is 
rooted to the star, while the dotted lines are connected to the surrounding disk.    
 \label{f3}}
\end{figure*}

\subsection{Behaviour of the solutions with $\Pi_*$}\label{subs43}

The effect on the solutions of increasing the pressure at the Alfv\'en distance $\Pi_* $ 
to values  $\Pi_* > \Pi_{*,\rm min}$ is 
shown in Fig. \ref{f2}, where $G(R)$, $\Pi(R)$ and $T(R)$ are
plotted for a positive and a negative value of $\epsilon$. 
First, we note that only for $\Pi_* \gg \Pi_{*,\rm min}$ the
solutions with positive and negative $\epsilon$ appear to have
similar profiles. And second, it is also evident that an increase of 
$\Pi_*$ always leads to a reduction of the jet radius, i.e. it has
basically the same effect as the increase of $\kappa$ (Fig. 1). 
This is expected because the pressure gradient across
the streamlines has terms which are $\propto \kappa \Pi$, 
Eq. (\ref{fnp}).  Nevertheless, the asymptotic Alfv\'en number 
and density are scarcely affected by the value of $\Pi_*$. 
And, the flow velocity ($\propto M^2/G^2$) increases with $\Pi_*$, 
a fact also related to an increase of the thermal driving efficiency.

\begin{table}
\caption{Parameters of the solutions  for $\lambda=3$ and $\delta=0.01$. $R_o$ and  $R_{cr}$
are given for $\Pi_{\star}=\Pi_{\star,\rm min}$.
\label{tab:tab1}}
\begin{tabular}{|c|c|c|c|c|c|c|}
\multispan 6\hfill $\epsilon/2 \lambda^2 = -0.1 $ \hfill \\ 
\hline $\kappa$ & $\Pi_{\star,\rm min}$ & $\Pi_{\star,max}$  & $R_o$  
&  $R_{cr}$ & $\nu$  \\
\hline 0      &  289 &   -   & 0.72  & 0.72 &  34.1  \\
       0.001  &  321 &  505  & 0.73  & 0.73 &  35.9  \\
       0.003  &  413 &  653  & 0.73  & 0.74 &  40.3  \\
       0.009  & 1951 & 2089  & 0.79  & 0.97 & 102  \\
\hline
\end{tabular}
\vskip 0.5 true cm
\begin{tabular}{|c|c|c|c|c|c|c|}
\multispan 6 \hfill $\epsilon/2 \lambda^2 = 0.25$  \hfill \\
\hline $\kappa$ & $\Pi_{\star,\rm min}$ & $\Pi_{\star,max}$  & $R_o$  
&  $R_{cr}$ & $\nu$  \\
\hline 0      &  214 &  -    & 0.60  & 0.62 & 24.1  \\
       0.001  &  236 & 2007  & 0.61  & 0.62 & 25.2  \\
       0.003  &  297 & 1135  & 0.61  & 0.63 & 28.2  \\
       0.009  & 1286 & 1423  & 0.45  & 0.88 & 65.1  \\
\hline
\end{tabular}
\end{table}

The pressure at the Alfv\'en surface  $\Pi_* $ cannot be 
however arbitrarily large. Namely, the asymptotic pressure 
$\Pi_{\infty}$ increases with $\Pi_*$, and it may even exceed 
the corresponding pressure values at $R=1$, but there is a 
sharp decrease of $\Pi$ at the position of the bump (with a
consequent lowering of the temperature). This minimum value of the 
pressure at the bump decreases by rising $\Pi_*$, and above a 
maximum value ($\Pi_{*,\rm max}$) the pressure becomes negative at 
the bump.  Hence, acceptable solutions exist only for 
$\Pi_{*,\rm min} \leq \Pi_* \leq \Pi_{*,\rm max}$.
In other words,  the range of acceptable values of $\Pi_*$ is quite
narrow, and further shrinks by decreasing $\epsilon/ 2 \lambda^2 $ 
and increasing $\kappa$ or $\delta$ (see Table 1). 
It turns out that below  a
threshold value of $\epsilon/2 \lambda^2$ the solutions are unphysical
because we obtain $\Pi_{*,\rm min} > \Pi_{*,\rm max}$, unless $\kappa
\rightarrow \delta$. For example for $\epsilon/2 \lambda^2 = -5$
we have acceptable solutions only for $\kappa=0.009$ with $4750
\leq \Pi_* \leq 4835$, besides the solution with $\kappa = 0$
and radially expanding streamlines.

Concerning the asymptotic velocities of the collimated outflow,
with the assumed parameters, as $\kappa = 0.001 \rightarrow
0.009$ and $\Pi_* = \Pi_{*,\rm min} \rightarrow \Pi_{*, \rm max}$  we find
for $\epsilon/2 \lambda^2=0.25$,  
$M = 3.5 \div 8$ and  $V_r/V_* = 5 \div 5 \times 10^2$, 
while for $\epsilon /2 \lambda^2=-0.1$ we get 
$M = 3 \div 15$ and $V_r/V_* =20 \div 4\times 10^2$ .

\subsection{Summary of the parametric analysis}\label{subs44}

\begin{figure}
\centerline{
\psfig{figure=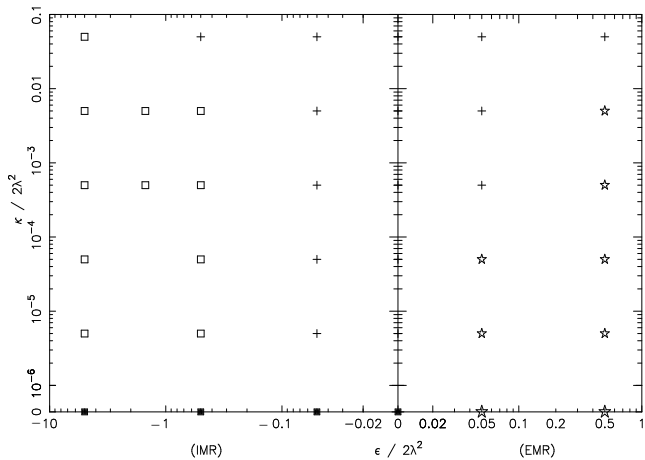,width=8.50truecm,angle=0}}
\caption{Summary of the various domains of the pressure behaviour
in the parameter space 
[$\epsilon/2\lambda^2$, $\kappa/2\lambda^2$] for $\nu=\lambda=1$. 
We show the domains of magnetic collimation
(stars), pressure confinement (crosses), the forbidden region
where all collimated solutions obtain at the bumps a negative 
pressure (open squares), and finally the domain of the radial 
solutions (filled squares).
\label{f4} }
\end{figure}

In order to summarize the trends we just 
discussed, we illustrate in Fig. \ref{f4} how the behaviour of the
solutions with the pressure is displayed in the parameter space
($\epsilon$, $\kappa$) for an arbitrary and illustrative set of 
values $\lambda = \nu = 1$.  
The exact location of the various domains in this space 
[$\epsilon$,$\kappa$] depends strongly on the set of the 
remaining parameters. 
The following discussion in each case refers to the solution 
corresponding to the minimum pressure $\Pi_{*,\rm min}$ allowed, 
except where it is otherwise indicated.

In the domain of Efficient Magnetic Rotators (EMR, $\epsilon>0$),
jets are cylindrically collimated by magnetic forces, as expected
(domain with stars in Fig. \ref{f4}), unless $\kappa$ is getting too 
large in which case we enter the regime of pressure confinement
(crosses in Fig. \ref{f4}). This pressure confined
region extends to the domain of Inefficient Magnetic Rotators (IMR) on
the left part of Fig. \ref{f4}. However, when $\epsilon$ is getting too 
negative, it occurs that $\Pi_{*,\rm min} > \Pi_{*,\rm max}$ for small enough values of $\kappa$
(open squares in Fig. \ref{f4}). This region  
corresponds in principle to physically non acceptable collimated solutions; 
we shall see however in the following subsection that a subdomain 
exists where radial solutions may exist.  

We remark that these main features of the numerical
solutions we just discussed are not qualitatively affected by
different values of the parameters. The main effect of lower
values of $\lambda$ is to shift downstream the `X' type critical
point closer to the Alfv\'en point (see ST94), and the properties
of the solutions appear very similar for equal values of the
quantity $(\delta - \kappa)\nu^2$.

In the previous subsections, we have seen that slight increase 
of the magnitude of $\kappa$ and/or $\Pi_*$ sharply reduces 
the jet radius (mainly for $\epsilon <0$). 
The jet radius may even asymptotically become smaller than 
at $R=1$ (i.e., $G_{\infty} < 1$).  This configuration
implies a reversal of the electric current somewhere in the
superAlfv\'enic region, and this may still be consistent with a
reasonable topological structure of the outflow (see e.g. VT98,
Vlahakis 1998). However jets with such a decreased transverse
size far from the central object are likely to be unrealistic.
Therefore we could assume in general that the parameters should be
constrained such that $G_{\infty} > 1$. We must note also that
when $G_{\infty} <1$ the asymptotic pressure is rather high (see
Fig. \ref{f2}) while it is reasonable to assume that in astrophysical
jets  $\Pi_{\infty} \ll \Pi_*$. From all these arguments it
appears that collimated solutions suitable to model collimated 
astrophysical outflows should be in general selected with 
$\Pi_* \gapp \Pi_{*, \rm min}$.

\section{Limiting solutions\label{sec5}}

\begin{figure}
\centerline{
\psfig{figure=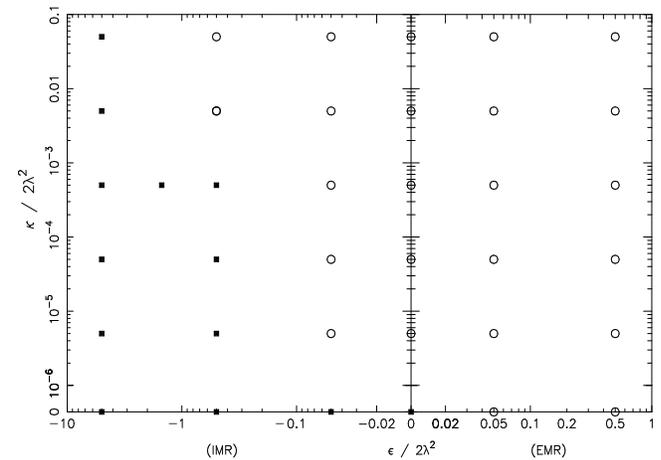,width=8.50truecm,angle=0}}
\caption{In the same parameter space as Fig. 4   
we show the limiting solutions between the collimated solutions and the
flaring ones which are either collimated without oscillations but
negative $\Pi_{\infty}$ (circles) or which are  radially
expanding at least up to $R=10^6$ with $\Pi_{\infty}=0$ (filled 
squares) 
\label{f5}}
\end{figure}

\begin{figure*}
\centerline{\psfig{figure=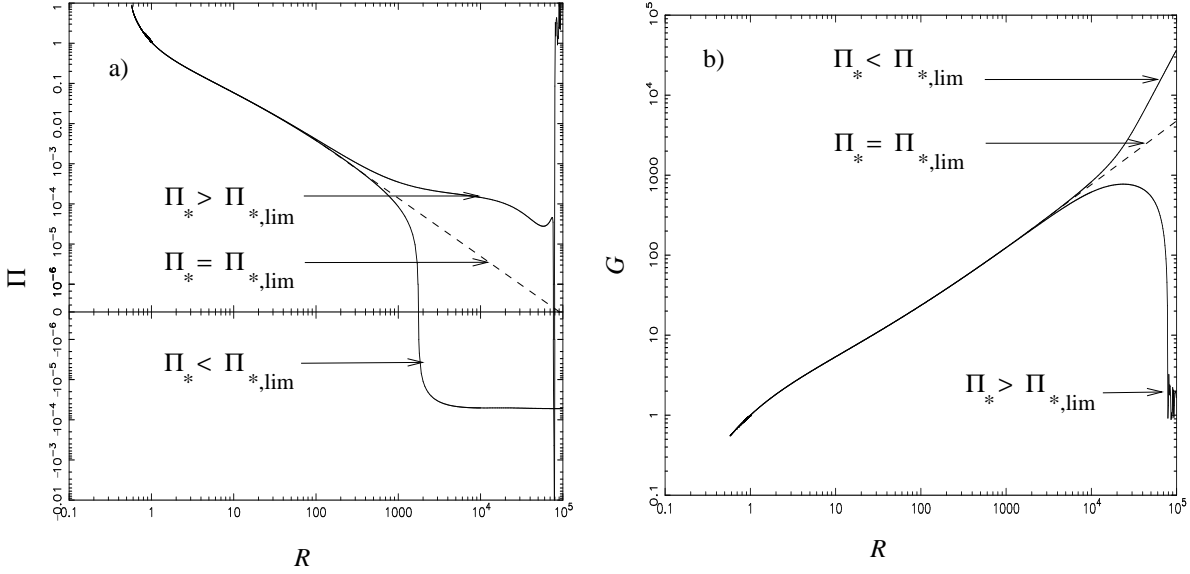,width=16.0truecm,angle=0}}
\caption{$\Pi (R)$ (panel a) and $G(R)$ (panel b) are plotted for  
$\epsilon/2 \lambda^2 =-0.5$, $\kappa=0.001$, $\nu=\lambda=1$. 
By tuning $\Pi_*$ to a special value $\Pi_* =\Pi_{*, \rm lim}$ 
we find, between a sharply recollimated ($\Pi_* > \Pi_{*, \rm lim}$) 
and a flaring solution ($\Pi_*<\Pi_{*, \rm lim}$), a limiting solution 
(dashed line) which is radial up to $R \approx 10^6$ (see the plot 
of the streamlines  in Fig. \ref{f8}a). 
\label{f6} }
\end{figure*}

\begin{figure*}
\centerline{\psfig{figure=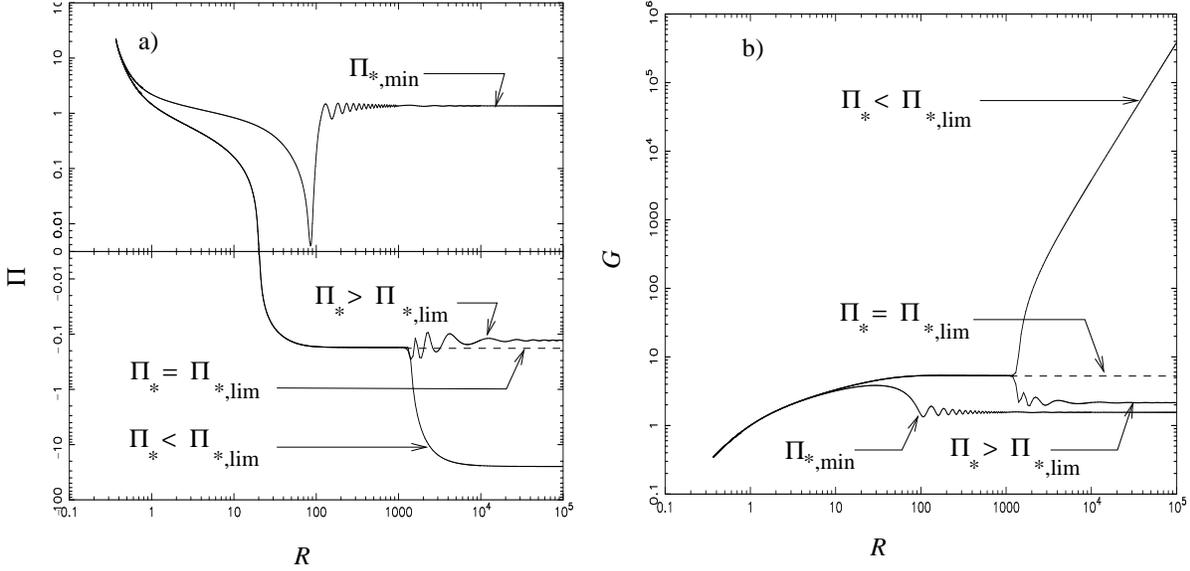,width=16.0truecm,angle=0}} 
\caption{The same as in Fig. 6 for $\epsilon=0$, 
$\kappa=0.01$, $\nu=\lambda=1$: the limiting solution now is collimated
without bumps. We plot also the solution with positive pressure everywhere,
labeled with  $\Pi_{*, \rm min}$. The streamlines of the limiting solution 
and of the one with $\Pi_* > \Pi_{*, \rm lim}$  are shown in  
Figs. \ref{f8}b and \ref{f8}c, respectively. 
\label{f7}}
\end{figure*}

We outline here the properties of two kinds of solutions (that we call 
`limiting' solutions) that are outside the boundary conditions on the 
pressure we discussed in the previous section.

\subsection {Radially expanding limiting solutions}
\begin{figure*}
\centerline{
\psfig{figure=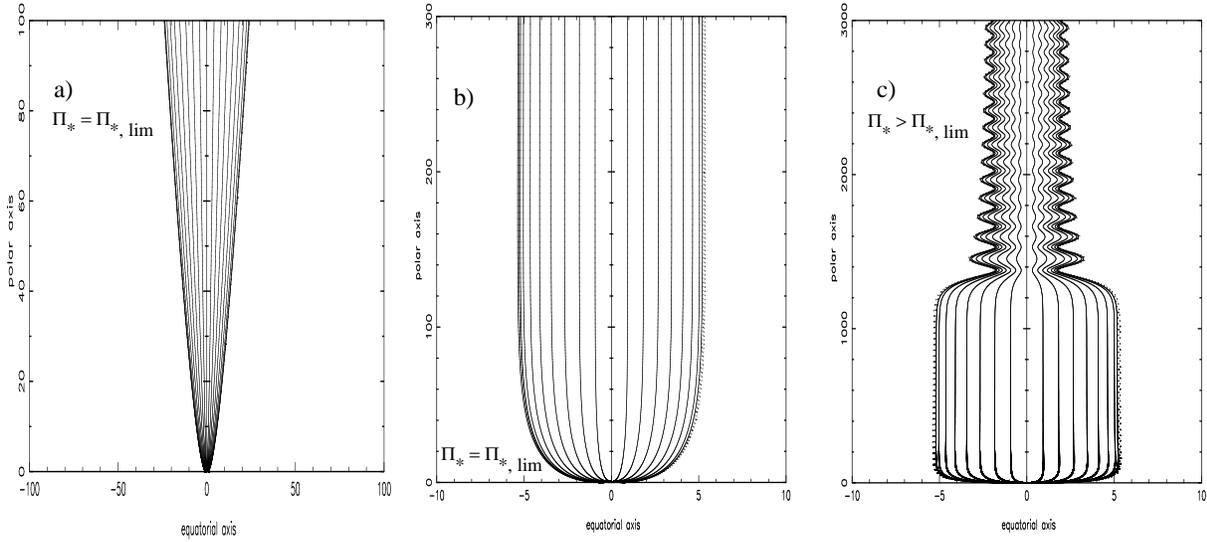,width=16.0truecm,angle=0}}
\caption{Sketch the poloidal streamlines of the radial limiting 
solution of  Fig. \ref{f6} (panel a). In panels (b) and (c) 
we show the poloidal streamlines of the limiting solution (collimating without
oscillations) and of the one  sharply recollimated with 
$\Pi_*>\Pi_{*, \rm lim}$, reported   
in Fig. \ref{f7}.  
\label{f8}}
\end{figure*}

We have seen that for sufficiently negative values of $\epsilon$ and
small values of $\kappa$, the sharp jet recollimation
leads to a negative pressure, with $\Pi_{*,\rm min}$ becoming
larger than $\Pi_{*,\rm max}$. These solutions have been rejected 
as unphysical (open squares in Fig. \ref{f4}). However it turns out 
that by further decreasing $\Pi_*$ 
below $\Pi_{*,\rm min}$ we find  a subdomain of this region
(filled squares in Fig. \ref{f5}), where we could 
follow numerically solutions that
remain radial up to $R=10^6$ by tuning
the initial pressure to a specific value $\Pi_*=\Pi_{*, \rm lim}$. 
An example of such a solution is given in Figs. \ref{f6} and  \ref{f8}a.  
These
solutions, with an almost vanishing pressure and radial asymptotics, 
are limiting solutions between sharply recollimated ones (for
$\Pi_*>\Pi_{*, \rm lim}$, see Fig. \ref{f6}) and flaring
solutions (for $\Pi_*<\Pi_{*, \rm lim}$).
Numerically we are unable to say if such solutions are strictly
radial asymptotically, or if they become cylindrical with
a negative pressure far away from the base, as we discuss 
hereafter. However
they can model radial winds up to the region of the outer
asteropause where the wind is shocked with the interstellar
medium because we do not expect the shock to be that far away
(i.e. farther than $R=10^6$). Thus they allow some extension of
the domain of radial winds into that of underpressured flows
and they are well adapted to model winds like the solar wind 
(Lima et al. 2001a).

\subsection {Non oscillating limiting solutions}

In the remaining region of Fig. 5 (circles) the limiting
solution between flaring and recollimating ones is a solution
that collimates into cylinders without oscillations or bumps, 
and with negative $\Pi_{\infty}$. An example is presented
in Figs. \ref{f7} and \ref{f8}b. This
result is not surprising and can be related to that 
obtained by TTS97, where two families of solutions were found for
fixed streamlines. The first one was composed of thermally confined
and underpressured jets showing the presence of a bump in the
pressure. The second family was composed of initially
underpressured jets which become overpressured downstream and
magnetically confined without bumps. We have the analogous result
here, at least in the domain of negative $\epsilon$.

But can this solution with negative $\Pi$ be acceptable? In
fact, the pressure can be defined modulus an arbitrary constant, 
\begin{equation}\label{pressureP0}
P(R,\alpha) = {1 \over 2} \rho_* V^2_* \Pi(R)[1+ \kappa \alpha] +
P_o \,.
\end{equation}
Thus we can drop the constraint $
\Pi_* \geq \Pi_{*,\rm min}$ by choosing values of this constant
such that the pressure is always positive. However the increase of
$P$ implies also higher temperatures of the plasma,
that should be eventually detectable with observations.
Furthermore, high gas pressure could require unreasonable large
heating processes in the plasma. So adding $P_o$ to
solutions that present a bump in the pressure (as in Fig. \ref{f8}c) 
is likely to be unrealistic. Conversely, including this 
constant to the limiting 
solutions with no bump (e.g. Fig. \ref{f8}b) is feasible in principle 
and should not require a much larger amount of heating. We could 
speculate that such solutions could be interesting to model jets 
which are initially underpressured
(in the region where $\Pi > 0$) and overpressured in the outer
zone (where $\Pi < 0$).

\section{Dynamics of the outflow\label{sec6}}

\subsection{Force balance along and across the flow\label{sec6.1}}

A complete physical understanding of the behaviour of the
solutions discussed in the previous section may be obtained by 
studying systematically the various forces acting on the outflow. 
In STT99 a similar analysis was attempted in asymptotically 
cylindrically collimated outflows. In such a case the
jet had attained its terminal speed and force balance was 
studied only in the normal to the fieldlines direction. 
In the present case we consider all forces across and along the flow, 
in the whole region, from the base to the asymptotic zone. 
For this purpose, the original MHD set of equations
can be projected on the poloidal plane and there they may be split 
in two directions, one being tangent ($\hat s$) and the other ($\hat n$)
perpendicular to a particular poloidal streamline 
$\alpha=$ const. In this local system of orthogonal coordinates, 
the most general form of the MHD equations reduces to:
\begin{eqnarray}
\label{sequation}
\parallel {\hat s}:\;\; 0 = - \rho V_p \frac{\partial V_p}{\partial
s} - {\partial  P\over  \partial s} -\rho {\partial  {\cal V}
\over \partial  s} + \left( \frac{\rho V^2_{\varphi}}{\varpi} -
\frac{B^2_{\varphi}}{4\pi\varpi}\right)
{ {\partial} \varpi \over {\partial} s} \nonumber\\  -
 { {\partial} \over  {\partial} s}\left(
\frac{B^2_{\varphi}}{8\pi} \right ) \,,\;\;\;\;\;\;
\end{eqnarray}
\begin{eqnarray}
\label{nequation}
\parallel {\hat n}:\;\; 0 = - \frac{\rho V_p^2}{R_c} 
- { {\partial} P\over  {\partial} n}
-\rho { {\partial} {\cal V} \over {\partial} n}
+ \left( \frac{\rho V^2_{\varphi}}{\varpi} -
\frac{B^2_{\varphi}}{4\pi\varpi}\right) { {\partial} \varpi \over
{\partial} n}   \nonumber\\
- { {\partial} \over  {\partial} n}\left(
\frac{B^2_{\varphi}}{8\pi} \right ) - { {\partial} \over
{\partial} n}\left( \frac{B^2_{p}}{8\pi} \right )
+\frac{B^2_p}{4\pi R_c} \,,\;\;\;
\end{eqnarray}
where $\varpi = r \, {\rm sin} \theta$, ${\cal V}$ is the gravitational 
potential ($=-{\cal G M}/r$)
and $R_c$ is the radius of curvature of the poloidal streamline. The
vectors ${\hat s}$ and ${\hat n}$ are directed towards
the outer regions and the polar axis, respectively. Then, a force
$\vec{f}$ has two components, $f^{\hat s}$ and $f^{\hat n}$.  A
positive component $f^{\hat s}$ accelerates the flow along the 
streamline, while a positive value for $f^{\hat n}$ collimates the
plasma. Conversely, negative $f^{\hat s}$ and $f^{\hat n}$ imply 
that the wind is decelerated and that the streamlines `flare' away  
from the axis, respectively. 

The full expressions of the forces are given and discussed in detail 
in Appendix C. 
In the following we outline the
main dynamical properties along and across the streamlines,
referring to the cases of collimated flows we have discussed  in
Sec. \ref{sec4} and plotted in Fig. \ref{f3}.

\subsection{Force balance along ${\hat s}$\label{6.2}}

\begin{figure*}
\centerline{
\psfig{figure=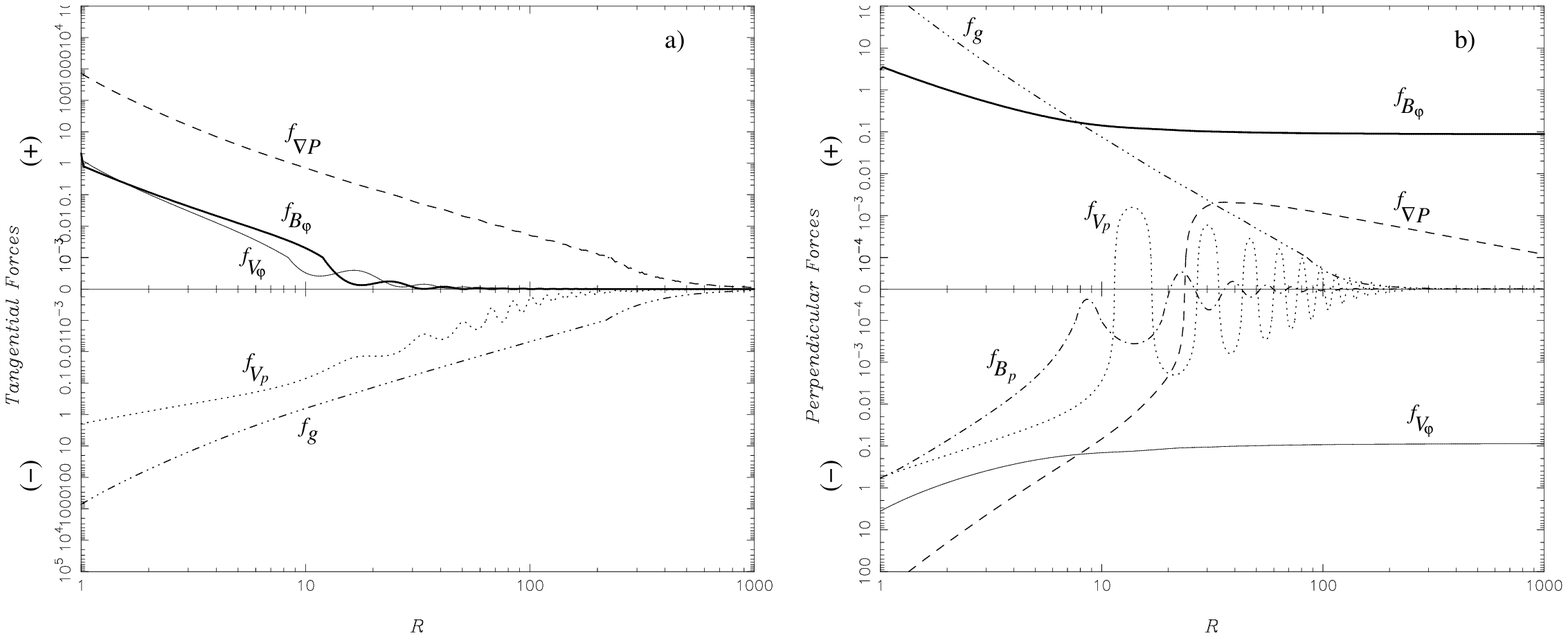,height=8.0truecm,width=16.0truecm,angle=0}}
\caption{Plot of the different forces acting along ($\hat s$,  left
panel) and perpendicular to the poloidal streamlines ($\hat n$, right
panel) for $\kappa = 0.003$, $\epsilon/2 \lambda^2=0.25$ and
$\Pi_* = \Pi_{\rm min}$ (Fig. \ref{f3}b). We plot the negative of the pressure gradient 
(dashes), the gravitational force 
(dash-three dots), the poloidal inertial and curvature forces (dots), 
the poloidal 
magnetic force (dash-dot), the centrifugal force (thin solid line) and 
the total toroidal 
magnetic force (thick solid line).
\label{f9}}
\end{figure*}

\begin{figure*}
\centerline{
\psfig{figure=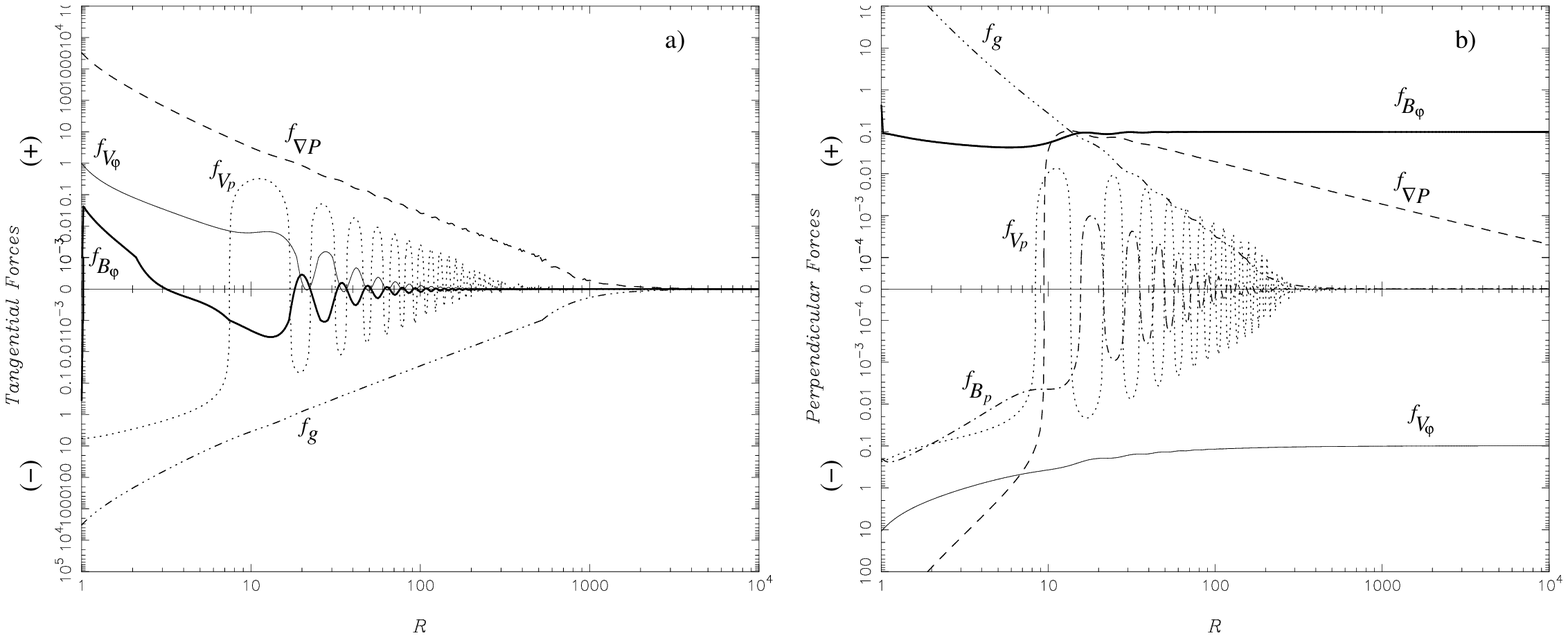,height=8.0truecm,width=16.0truecm,angle=0}}
\caption{The same as Fig. \ref{f9} for $\kappa = 0.009$, $\epsilon/2 \lambda^2=0.25$ and
$\Pi_* = \Pi_{\rm min}$ (Fig. \ref{f3}d).
\label{f10}}
\end{figure*}

It is easy to see that, the centrifugal force and the negative
gas pressure 
gradient for a monotonically decreasing $\Pi(R)$, always
accelerate the flow (see left panels of Figs. \ref{f9} - \ref{f12}). 
On the other hand, gravity and the hoop stress associated with the toroidal 
magnetic field always decelerate the plasma. 
However the initially negative
gradient of $B^2_{\varphi}/8\pi$ dominates close to the base and accelerates 
the flow (see left panels of Figs. \ref{f9} - \ref{f12}).  
Finally, as is well known, the poloidal component of the magnetic field has 
no effect  on the plasma dynamics along $\hat s$. 
From the left panels of Figs. \ref{f9} - \ref{f12},   we see that 
the outflow is always thermally driven, with the magnetic and 
centrifugal forces playing a rather marginal role in the acceleration.  
For the assumed parameters the overall acceleration occurs 
for $R \simless 10^3$, independently of the values of $\epsilon$ 
and $\kappa$. Oscillations in the flow variables may exist in
some cases. For example, when $\epsilon < 0$ 
and for intermediate values of $\kappa$ (Fig. \ref{f11}), 
asymptotically  the pressure
gradient reacts to the oscillations of the streamlines, as
expected in thermally confined jets (STT99). 
For larger values of $\kappa$ the flow appears 
already collimated in the acceleration
region (see below), so the oscillations appear much closer to the
Alfv\'enic surface (Figs. \ref{f10} and \ref{f12}, left panels).

\noindent
\noindent

\subsection{Force balance along ${\hat n}$\label{6.3}}

The analysis of the forces acting perpendicularly to a poloidal
streamline is more complicated. First, the centrifugal force is always
negative tending to open the streamlines. Second, both the tension
and the gradient of the toroidal magnetic field are directed
towards the axis, assisting collimation. Also, the normal component of the
gravitational field tends to collimate the flow when the streamlines 
do not flare faster than the spherical expansion. 
Two terms contribute to the gradient of the pressure,
with opposite effects: the term related to the radial component
($\propto {\rm d} \Pi / {\rm d} R$) decollimates the fieldlines
while the one related to the meridional component ($\propto \kappa
\Pi$) is positive and confines the flow. Concerning finally the
poloidal magnetic field, its gradient is negative (decollimating) 
while its tension is positive (collimating). 

\begin{figure*}
\centerline{
\psfig{figure=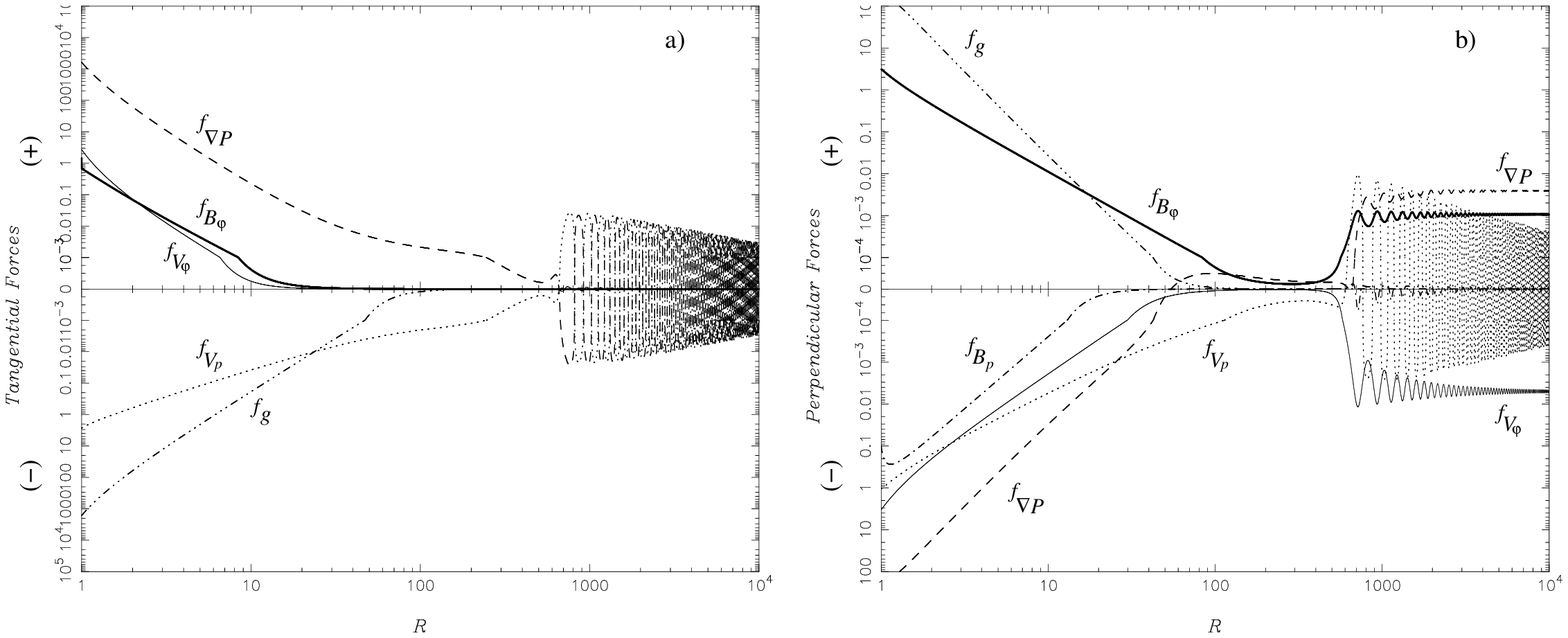,height=8.0truecm,width=16.0truecm,angle=0}}
\caption{Plot of the different forces acting along ($\hat s$,  left
panel) and perpendicular to the poloidal streamlines ($\hat n$,  right
panel) for $\kappa = 0.003$, $\epsilon/2 \lambda^2=-0.1$ and
$\Pi_* = \Pi_{\rm min}$ (Fig. \ref{f3}a). As in Fig. \ref{f9} we plot the negative of the 
pressure gradient (dashes), the gravitational force 
(dash-three dots), the poloidal inertial and curvature forces (dots), the poloidal 
magnetic force (dash-dot), the centrifugal force (thin solid line) and the total toroidal 
magnetic force (thick solid line).
\label{f11}}
\end{figure*}

\begin{figure*}
\centerline{
\psfig{figure=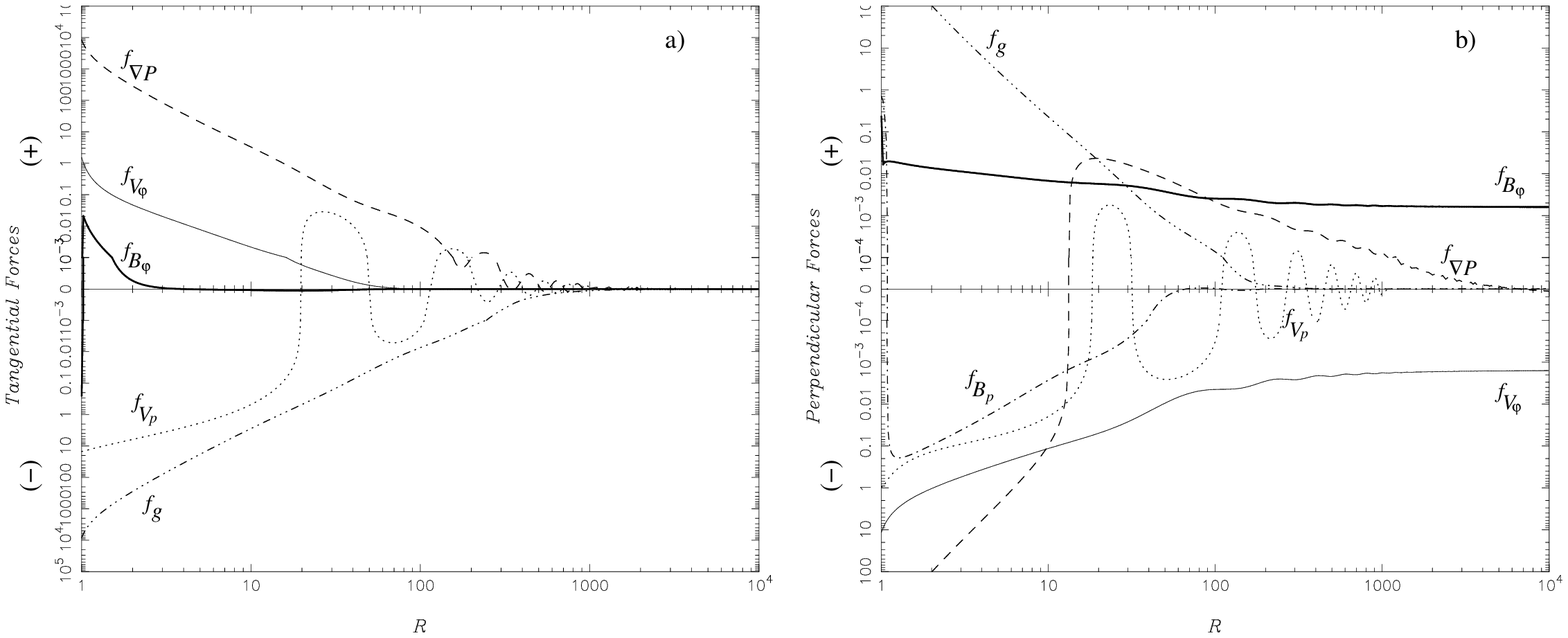,height=8.0truecm,width=16.0truecm,angle=0}}
\caption{The same as Fig. \ref{f11} for $\kappa = 0.009$, $\epsilon/2 
\lambda^2=-0.1$ and $\Pi_* = \Pi_{\rm min}$ (Fig. \ref{f3}c).
\label{f12}}
\end{figure*}

The interplay of all these forces is shown in the right panels of
Figs. \ref{f9} - \ref{f12}. For positive $\epsilon$ and
intermediate $\kappa$ (Fig. \ref{f9}) the dynamics of
the streamlines in the subAlfv\'enic and transAlfv\'enic region
($R \simless 10$) is governed by gravity and the radial
component of the pressure gradient. Farther away the magnetic and
centrifugal forces prevail: the centrifugal force is  balanced by
the hoop stress and the gradient of the toroidal magnetic field
(these two components provide the same contribution to this
force). Asymptotically then the flow is magnetically confined (see
also STT99). The effect of the other forces is quite negligible:
they just show an oscillating behaviour that slightly affects the
whole structure. This same picture basically holds if we increase the
value of $\kappa$, with an increased effect of the thermal pressure
(Fig. \ref{f10}, right panel). Namely,  the  wind is always
asymptotically magnetically collimated but in an intermediate
region ($R \approx 15$) the outflow is thermally confined.
Oscillations are always present in the subAlfv\'enic and the
intermediate regions with larger amplitude: this is likely to be related
to the larger effect of the pressure on a more collimated wind in
this intermediate region.

A similar scenario is also found for negative $\epsilon$ and large
values of $\kappa$ (see Fig. \ref{f12}, right panel). The  wind is also
asymptotically magnetically collimated but with a larger  intermediate
region ($R \approx 20 \div 100$) where the outflow is thermally confined. 

For smaller
$\kappa$ (Fig. \ref{f11}, right panel), differently from the case
with positive $\epsilon$, the toroidal magnetic force balances the
curvature force in an intermediate  region ($R \approx 10 \div 100$) 
but without any confinement of the flow that
expands almost radially in this zone. The collimation starts
downwind where the gas pressure prevails on the magnetic force
($R \approx 100 \div 400$),
leading to a cylindrical configuration with oscillations 
(see Fig. \ref{f3}a). In this
asymptotic region the centrifugal force is balanced by the
gradient of the pressure, with a small contribution from the
toroidal magnetic field, since we are there in the thermally 
confined regime.


\section{Summary and astrophysical implications \label{sec7}}

In this paper we confined our attention to the study of outflows 
with a density increasing from the axis towards the 
surrounding streamlines [cf. Eq.  (\ref{density}) with $\delta >0$] 
faster than the pressure does [cf. Eq.  (\ref{pressure}) 
with $\kappa >0$], i.e. $\delta > \kappa$.
 
In such {\it underpressured outflows}, the temperature is peaked at the axis 
relatively to the surrounding regions.  Such centrally hotter 
astrophysical atmospheres are likely to exist in coronae wherein, for
example, strong 
Alfv\'en waves propagating along the direction of the polar magnetic 
field deposit enough energy which raises thus the plasma temperature. 
The inevitable consequence of this high temperature is an induced 
wind-type outflow,  which is dominantly ``thermally'' driven around the 
rotation axis where magnetocentrifugal forces are negligible (see Figs. 
\ref{f9} 
- \ref{f12} ). By thermally we mean not only thermal conduction but also 
waves, etc, which can be effectively included in the pressure gradient term.

After the first acceleration stage where the gas expands almost radially 
up to the transAlfv\'enic region, if the outflow carries an electrical current
and the magnetic rotator is sufficiently efficient (EMR), 
it is collimated magnetically via the combination of the hoop stress and 
the pressure gradient of the azimuthal magnetic field in the superAlfv\'enic
regime.  
Several jets from Young Stellar Objects have probably such smoothly increasing 
radius with a large terminal transverse extension. They would naturally 
correspond to
the magnetic collimation produced by an EMR. Outflows from an EMR may also 
describe jets from
Radio Loud AGNs like FR Is and FR IIs (see Sauty et al. 2001 for a possible 
application of the
present solutions to AGNs). For definite conclusions in this last case, 
however, 
a relativistic extension of our model is needed.

On the other hand, the produced outflow may also be collimated via 
thermal pressure gradients, if it is underpressured on the axis. 
Such pressure confined jets tend to show a rather strong recollimation. 
This could be related with some jets from slowly rotating T Tauri stars and
Planetary Nebulae which show a rather strong recollimation, or some choked
winds from Seyfert Galaxies.  For example, 
the peculiar jet of RY Tauri originates in a very slowly
rotating CTTS and seems to recollimate at 38 stellar radii from
the base (see Gomez de Castro et al. 2001).
The sharp gradient of the gas pressure in  the recollimating
region could lead naturally to the formation of a shock which may be
consistent with the scenario of recollimation by internal shocks
proposed in the literature. 

In very Inefficient Magnetic Rotators (IMR) like the sun, radial solutions as 
the one plotted in Fig. \ref{f6} are likely to apply to  winds produced 
by such 
sources (Lima et al. 2001b).  
Thus, jets from Young Stars may evolve from a narrowly collimated magnetic
outflow to a radially expanding wind. Such radial solutions could also 
correspond to
winds from Seyfert Galaxies.

If we were to apply recollimating  solutions to jets from
Planetary Nebulae, we would be inclined to think that in these
objects recollimation is an effect of the pressure gradient rather
than a pure magnetic pinching. Nevertheless, the jet could be
ultimately magnetically confined once it has refocalized.
 
Interesting solutions for underpressured jets becoming
overpressured after exiting from the central embedding medium can be 
found. They correspond to the limiting solution plotted in Figs. \ref{f7}
and \ref{f8}b. Such outflows are magnetically confined and
in contrast with the previous ones do not exhibit any recollimation or
oscillations. They may describe Young Stellar jets exiting from the 
central cloud.

More precise modelling of various specific astrophysical objects is 
underway and will be presented elsewhere; here we just 
summarize some of the results obtained so far.
Preliminary applications to the solar wind are given in Lima et al. 
(2001b). By analysing Ulysses data we arrived at the following ranges  of 
the 
model parameters for the solar wind : $\delta\approx 2- 4$, 
$\lambda\approx 0.1$, $\kappa\approx 0.3 - 0.8$ and  $\nu\approx 0.7$. 
With these parameters a reasonable fitting of all observed 
quantities is obtained, except perhaps the magnetic field at large distances,
 which turns 
out to be a bit too weak compared to the measured one. 
The above values of the parameters imply $\epsilon/2\lambda^2 \sim -50$, 
thus  confirming the strong inefficiency of the solar magnetic rotator. 
In such a case the solutions are expected to be qualitatively similar to those 
shown in Fig. 3a, i.e. radially expanding streamlines connected through the 
bump to the far dowstream collimated region. In the present case, with $\kappa 
\ll \delta$ and very low asymptotic pressure, the position of the bump is 
very far away from the Alfv\'enic  surface, then the streamlines are expected 
to be basically radial in the region explored by the spacecraft.

Parallel to this, preliminary results for jets from T Tauri stars 
(Meliani 2001) provide the following range of the values of the parameters: $\delta 
\approx 0.1 - 0.2$, $\lambda\approx 0.8$, $\kappa\approx 0.01 - 0.06$, 
$\nu\approx 1 - 6$ and $r_*\approx 10 \; r_\star \approx 20-30 \; r_\odot$. 
It is worth noting that observations of these objects imply the existence of 
a UV emitting region near the central star, which may possibly be
associated with a shock 
(e.g. in RY Tau this UV zone is at $\approx$ 38 $r_\star$).
With the above parameters we deduce $\epsilon/2\lambda^2\approx 0$  
(negative or positive), which means that T Tauri stars correspond to more 
efficient magnetic rotators. 
For the above ranges of the parameters, the solutions are expected to
be similar to those reported in Fig. 7 and 8. In particular we could have a 
recollimated solution with the UV shock at the position of the bump, or 
a limiting solution with monotonically collimated streamlines without
oscillations (see Fig. 8b), in which case the shock would result from the 
growth of some internal instability. Both types of solution reproduce 
in a satisfactory way the  velocity and density profiles, the jet 
morphology and the rotational rate. Yet, the second type of solutions leads 
to lower temperature profiles which suggests that a thermally driven wind 
could account for the observed data in such objects, even without extra 
nonthermal heating. 

To distinguish between recollimated and limiting solutions, we may ask about 
the stability of our solutions. 
Some hints about this crucial point can be given for the asymptotic 
region. By a local analysis using the thin flux tube approximation, Hanasz  
et al. (2000) have shown that the internal part of jets with rotational 
laws comparable to the one used in the present model (Eq.\ref{LOM}) are more 
stable to 
magnetorotational instabilities than the surrounding jet coming from the disk. 
In some cases rotation completely suppress the instability.
Of course, we do not claim that this analysis is complete but it may be an 
indication that azimuthal magnetic fields do not necessarily destroy the 
global structure of MHD outflows modelled by our solutions.

However, while the properties of the local stability of 
cylindrical jets are quite well known in the linear regime and for simple 
equilibrium structures (see e.g. Birkinshaw 1991, Ferrari 1998, and references
therein, Kim \& Ostriker 2000),  unfortunately a global, non-linear analysis of the 
stability of 
general axisymmetric MHD steady flows is not yet available. 
This still remains a challenge for future studies. Furthermore, the present solutions 
may be used for the testing and interpretation of numerical codes employed for 
such a stability analysis.

To conclude with, let us stress that in the present study cylindrical jets 
\underbar{and} radial winds can carry a net current,
at least in the central part close to the axis, because they are
not necessarily force-free asymptotically. This is not in
contradiction with the main conclusion in Heyvaerts \& Norman (1989), 
because some of the
approximations done there do not necessarily hold close to the
axis. Our description is also only valid at the central part of the 
outflow, and  the current should close either in the surrounding wind, the external 
medium,  or in a current-carrying sheet around the jet.

\begin{acknowledgements}
E.T.  acknowledges financial support from the Observatoire de Paris, from the 
University of Crete and from the Conferenza dei Rettori delle Universit\`a 
Italiane (program Galileo). 
C.S. and K.T. acknowledge financial support from the French Foreign Office 
and the  Greek General Secretariat for Research and Technology (Program Platon and Galileo).  
\end{acknowledgements}

\appendix

\section{Generalities on self-similar solutions}\label{A.1}

For the sake of generality the free functions $A(\alpha)$,
$\Psi_A(\alpha)$ and $\Omega(\alpha)$, can be defined  through the
following new functions (for details see VT98):
\begin{eqnarray}
g_1(\alpha) \propto \int {\cal F}_1(A^{\prime}) {\rm d} \alpha,\;\;\;\;\;
g_2(\alpha) \propto  \int \Omega^2 \Psi^2_A {\rm d} \alpha, \nonumber \\
g_3(\alpha) \propto {\cal F}_2(\Psi_A)
\,,
\end{eqnarray}
\noindent
where $A^{\prime} \equiv {\rm d} A/ {\rm d} \alpha$. We remark that $g_1$ is
 related to the magnetic field structure through the magnetic flux function,
 while $g_2$ and $g_3$ define the current and density distribution,
respectively. Suitable
 choices of ${\cal F}_1$ and ${\cal F}_2$ allow to select various classes
 of self-similar solutions.

\noindent {\underbar{\it Radially self-similar models}}. If we
assume $M \equiv M(\theta)$ (i.e. the surfaces with the same
Alfv\'en number are conical),  $G \equiv G(\theta)$ 
and in Eqs. (A.1):
\begin{equation}
{\cal F}_1(A^{\prime}) = {{A^{\prime 2}} \over \alpha},\;\;\;\;
{\cal F}_2(\Psi_A) = \int {{\Psi^2_A} \over {\alpha^{3/2}}} {\rm d}
\alpha
\,,
\end{equation}
\noindent
we obtain a class of solutions that are self-similar in the radial
direction, used to model outflows from disks. Assuming for instance $g_1, g_2,
g_3 \propto \alpha^x/(x-2)$ (see Table 3 in VT98) with $x=3/4$ we have the well
known solutions of Blandford \& Payne (1982).

\noindent {\underbar{\it Meridionally self-similar models}}. By
choosing $G \equiv G(r)$, $M \equiv M(r)$ (i.e. the surfaces with
the same Alfv\'en number are spherical) and:
\begin{equation}
{\cal F}_1(A^{\prime}) = A^{\prime 2},\;\;\;\;
{\cal F}_2(\Psi_A)  = \Psi^2_A
\,,
\end{equation}
\noindent
we obtain the class of $\theta-$ self-similar solutions. If we
assume as the simplest case $g_1=\alpha$, $g_2= \lambda^2 \alpha$ and $g_3=1 +
\delta \alpha$ ($\lambda$ and $\delta$ constants, see Table 1 in VT98 ) we get
the solutions presented in ST94, TT97, STT99 and in the present paper.

\section{Equations for meridional self-similar flows)}\label{A.2}

From the definitions of Sec. \ref{sec22} we deduce the following expressions 
for the three components of the velocity and magnetic field (for details 
see ST94):

\begin{equation}
\label{Br}
B_r =B_{*} {1\over G^2(R)}\cos\theta\,,
\end{equation}
\begin{equation}
B_\theta =-B_{*} {1\over G^2(R)}{F(R)\over 2}\sin\theta
\,,
\end{equation}
\begin{equation}
\label{Bphi}
B_\varphi = - B_{*} {\lambda \over G^2(R)}
{\displaystyle 1 - G^2(R) \over 1 - M^2(R) }{R\sin\theta}
\,,
\end{equation}
\begin{equation}
\label{Vr}
V_r = V_{*}  {M^2(R)\over G^2(R)} { \cos\theta \over
\sqrt{1+\delta \alpha(R,\theta)}  },\;\;\;\;\;\;
\end{equation}
\begin{equation}
V_\theta =-V_{*} {M^2(R)\over G^2(R)}{F(R)\over 2}
{  \sin\theta \over \sqrt{1+\delta \alpha(R,\theta)}  }
\,,
\end{equation}
\begin{equation}
\label{Vphi}
V_\varphi = V_{*}  {\lambda \over G^2(R)}
{ G^2(R) - M^2(R) \over 1- M^2(R)}
{R\sin\theta \over \sqrt{1+ \delta \alpha(R,\theta) } }
\,.
\end{equation}

The three ordinary differential equations for $\Pi(R)$, $M^2(R)$ and
$F(R)$ are:
\begin{equation}\label{Eq1}
{\hbox {d} \Pi \over \hbox {d} R}=
- {2 \over G^4 }
   \left[ {\hbox{d} M^2 \over \hbox{d} R} + {M^2 \over R^2} (F-2) \right]
-  {\nu^2 \over M^2 R^2 }
\,,
\end{equation}
\begin{equation}\label{Eq2}
{\hbox {d} F(R) \over \hbox {d} R}={{{\cal N}_F(R,G,F,M^2,\Pi; \kappa, \delta,
\nu, \lambda)} \over \     {R \, {\cal D}(R,G,F,M^2; \kappa, \lambda)}}
\,,
\end{equation}
\begin{equation}\label{Eq3}
{\hbox {d} M^2(R) \over \hbox {d} R}={{{\cal N}_M(R,G,F,M^2,\Pi;
\kappa, \delta, \nu, \lambda)} \over \     {R \, {\cal D}(R,G,F,M^2; \kappa,
\lambda)}}
\,,
\end{equation}
where we have defined:
\begin{equation}\label{Eq4}
{\cal D}= (M^2-1)\left( 1+\kappa {R^2\over G^2} \right)
          + {F^2\over 4} + R^2\lambda^2{N_B^2\over D^2}
\,,
\end{equation}
\begin{eqnarray}\label{Eq5}
{\cal N}_F = -(\delta-\kappa)\nu^2 {R G^2\over 2 M^2}F
\nonumber\\
+ \left({2\kappa \Pi G^2 R^2} + (F+1)(F-2) \right)
\times
\nonumber\\
\times
\left(1+\kappa {R^2\over G^2} - {F^2\over 4}
                       -R^2\lambda^2{N_B^2\over D^3} \right)
\nonumber\\
+{M^2F\over4}(F-2)\left(F+2+2\kappa{R^2\over G^2}
             +2R^2\lambda^2{N_B^2\over D^3} \right)
\nonumber\\
-\lambda^2 R^2 F(F-2){N_B\over D^2}
\nonumber\\
+\lambda^2 R^2 \left(1+\kappa {R^2\over G^2} -R^2\lambda^2{N_B^2\over D^3}
                     - {F\over 2} \right)
\nonumber\\
\left( 4{N_B^2\over D^2}-{2\over  M^2}{N_V^2\over D^2}\right) \,,
\end{eqnarray}
\begin{eqnarray}\label{Eq6}
{\cal N}_M= (\delta-\kappa)\nu^2 {R G^2\over 2 M^2}(M^2-1)
\nonumber\\
+ \kappa \Pi R^2 G^2 M^2{F\over 2}
-{M^4\over 4}(F-2)(4\kappa {R^2\over G^2} +F+4)
\nonumber\\
+{M^2\over 8}(F-2)(8\kappa {R^2\over G^2} +F^2+4F+8)
\nonumber\\
- \lambda^2 R^2 (F-2){N_B\over D}
\nonumber\\
+\lambda^2 R^2 (2M^2+F-2)\left(
 {N_B^2\over D^2} -{1\over 2 M^2}{N_V^2\over D^2}\right)
\,.
\end{eqnarray}
with
\begin{equation}\label{Eq7}
N_B = 1 - G^2,\;\;\;\;N_V=M^2-G^2,\;\;\;\;D=1-M^2
\,.
\end{equation}
The meaning of the various parameters is discussed in Sec. 2.

At the Alfv\'en radius, the slope
of $M^2(R=1)$ is $p= (2 - F_*)/ \tau$, where $\tau$ is a solution of the
third degree polynomial:
\begin{equation}\label{Eq8}
\tau^3 + 2 \tau^2 + \left [ {{\kappa \Pi_*} \over {\lambda^2}} +
{{F^2_* - 4} \over {4 \lambda^2}} - 1 \right ] \tau +
{{(F_*-2)F_*} \over {2 \lambda^2}} = 0
\,,
\end{equation}
and  the star indicates values at $R=1$ (for details see ST94).


\section{Forces acting on the outflow on the poloidal plane}\label{A.3}
The components of the various MHD forces in the poloidal 
plane along the natural unit vectors ($\hat n, \hat s$) are related to 
those along the spherical coordinates poloidal unit vectors
($\hat r, \hat \theta$) via the following expressions:
\begin{equation}
f^{\hat s} (R, \theta ) = f^{\hat r} \cos \chi (R, \theta ) + f^{\hat \theta} \sin \chi
(R, \theta )
\,,
\end{equation}

\begin{equation}
f^{\hat n} (R, \theta ) =  f^{\hat r} \sin \chi (R, \theta ) - f^{\hat \theta} \cos \chi
(R, \theta )
\,,
\end{equation}

\begin{figure}
\centerline{
\psfig{figure=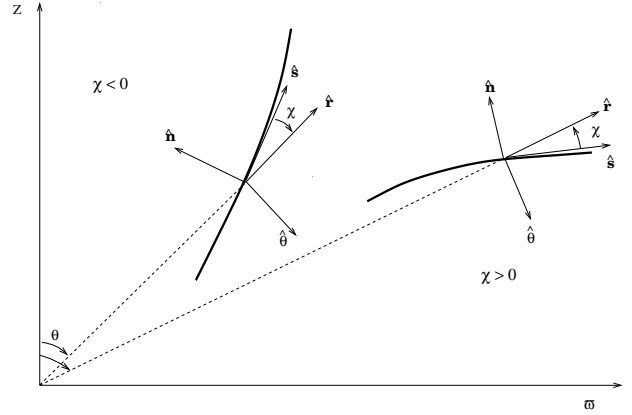,width=9.0truecm,angle=0}
}
\caption{Sketch of poloidal unit vectors for a converging ($\chi <0$) and a flaring  
($\chi >0$) streamline (thick line).
\label{f13}}
\end{figure} 
with  $\chi$ the angle a poloidal streamline makes with the radial
direction $\hat r$ (Fig. \ref{f13}). This angle is related to the expansion 
factor $F(R)$ and the polar angle
$\theta$ as,
\begin{equation}
\tan \chi (R, \theta )= - \tan \theta \frac{F(R)}{2}
\,. 
\end{equation}
When $\chi<0$  the poloidal streamline turns towards
the symmetry axis, while when $\chi >0$ it flares faster than radiality 
(see Fig. \ref{f13}).
By taking into account Eqs. (B.1) - (B.6) for the dependence of the 
physical quantities on $r$ and $\theta$, we obtain the 
following expressions for the meridional components 
of the various MHD forces $f^{\hat s}$ and $f^{\hat n}$ 
plotted in Figs.  \ref{f9} - \ref{f12}.  Note that all forces have 
the dimensions of $\rho_*V_*^2/2=B_*^2/4\pi$, a common factor 
which for simplicity we shall omit and the forces will 
thus be given dimensionless in the following (the different
forces and their direction are listed in Table C1).  

\noindent$\bullet$
{\it Gravity}: $ {\vec f}_g \equiv - \rho {\cal G M}  /R^2\, \hat r $. 

The two components of the gravitational force are:
\begin{equation}
f_g^{\hat s} \equiv - \rho { {\partial} {\cal V}\over  {\partial} s}
= -\cos \chi (R, \alpha )
{\nu^2 \over M^2 R^2 }(1+\delta \alpha )
\,,
\label{fsg}
\end{equation}
\begin{equation}
\label{fng}
f_g^{\hat n}\equiv - \rho { {\partial} {\cal V}\over  {\partial} n}
=-\sin \chi (R, \alpha )
{\nu^2 \over M^2 R^2 }(1+\delta \alpha )
\,.
\end{equation}
Note that always $f^{\hat s}_g < 0$, so that gravity in 
all cases acts to decelerate the flow, as expected.  
Conversely, $f^{\hat n}_g$ is positive and assists further  
a streamline which is turning toward the axis ($\chi < 0$), while it is
negative  assisting further decollimation when $\chi > 0$ and there is a 
flaring of the streamlines.  

\noindent$\bullet$
{\it Pressure gradient force}: $ {\vec f}_{\vec \nabla P}
\equiv - \nabla P(R, \alpha)$.

The two components of the pressure gradient along and
perpendicular to a poloidal streamline are:
\begin{equation}\label{fsp}
f_{\nabla P}^{\hat s}\equiv - { {\partial} P\over  {\partial} s} =
 -\cos \chi (R, \alpha )
{ {\rm d} \Pi \over  {\rm d} R} (1 + \kappa \alpha )
\,,
\end{equation}

\begin{eqnarray}\label{fnp}
f_{\nabla P}^{\hat n}\equiv - { {\partial} P\over  {\partial} n}
= - \sin \chi (R, \alpha ) \left\{
\left [
{\hbox {d} \Pi \over \hbox {d} R}  + \frac{4\kappa R \Pi}{FG^2}
\right]\right. \nonumber\\
+ \kappa \alpha
\left.\left[
{ {\rm d} \Pi \over  {\rm d} R}  + \frac{4\Pi}{FR}\left( \frac{F^2}{4} -
1 \right)
\right]
\right\}
\,.
\end{eqnarray}
\noindent
Note that $f_{\nabla P}^{\hat s}$ is always positive driving thus the wind 
whenever $\Pi$ decreases monotonically with $R$, as for 
example close to the hot base of the outflow (Figs. \ref{f9} - \ref{f12}, left panels).  
In the strongly oscillating regions $f_{\nabla P}^{\hat s}$ 
periodically accelerates  and decelerates the wind. 
The behaviour of the perpendicular component  $f_{\nabla P}^{\hat n}$ 
is more complex because it depends also on  
the strength of $\kappa \Pi $.  In the transAlfv\'enic 
region the pressure gradient is large, so that:

\begin{equation}
\label{fnp1}
f_{\nabla P}^{\hat n}  \approx - \sin \chi (R, \alpha ) (1 + \kappa \alpha)
{ {\rm d} \Pi \over  {\rm d} R}
\,,
\end{equation}
\noindent
while for $R \gg 1$, ${\rm d} \Pi /  {\rm d} R \rightarrow 0$ and

\begin{equation}
\label{fnp2}
f_{\nabla P}^{\hat n}  \approx - \frac{4\kappa R }{FG^2} \Pi \sin \chi (R, \alpha )
\,. 
\end{equation}
Hence, it becomes evident why in the subAlfv\'enic ($R < 1$) and 
transAlfv\'enic ($R\gapp 1$) regions the gradient of the gas 
pressure acts to `open' the fieldlines ($f_{\nabla P}^{\hat n} <0$), 
while further away ($R\gg 1$) the term $\propto \kappa \Pi$ 
acts to assist collimation for $\kappa >0$  ($f_{\nabla P}^{\hat n} >0$).

\begin{table*}
{\caption{Forces acting on a converging ($\chi < 0$) and a flaring 
($\chi > 0$) streamline: they are parallel or antiparallel to the unit
vectors $\hat s$ and $\hat n$ when they are $> 0$ or $< 0$, respectively. 
Then positive or negative $f^{\hat s}$ means that it accelerates 
or decelerates the flow, while positive or negative 
$f^{\hat n}$ means that it collimates or decollimates. Conversely $f_{V_p}^{\hat s} <0$
corresponds to acceleration and $f_{V_p}^{\hat s} >0$ to deceleration. 
\label{tab:tab2}}}
\begin{tabular}{|l|c|c|c|c|c|c|c|}
\hline
Force & Components & Conv. str. ($\chi < 0$) & Flar. str. ($\chi > 0$)  \\

\hline

Gravity    &  $f_g^{\hat s} \equiv - \rho {{{\partial} {\cal V}} \over {{\partial} s}}$
           &  $< 0$  & $ < 0$  \\
           &  $f_g^{\hat n} \equiv - \rho {{{\partial} {\cal V}} \over  {{\partial} n}}$
           &  $ > 0$ & $ < 0$  \\
\hline
Pressure Gradient  & $f_{\nabla P}^{\hat s} \equiv - {{{\partial} P} \over {{\partial} s}}$ & $ > 0$ & $ > 0$  \\
                   & $f_{\nabla P}^{\hat n} \equiv - {{{\partial} P} \over {{\partial} n}}$
                     & $ < 0$ $(|{\rm d} \Pi/{\rm d} R| \gg 0)$ & $> 0$ 
                       $(|{\rm d} \Pi/{\rm d} R| \gg 0)$      \\
                   &     &    $ > 0$ $({\rm d} \Pi/{\rm d} R \rightarrow 0)$ &
                       $ < 0$  $({\rm d} \Pi/{\rm d} R \rightarrow 0)$          \\
\hline
Centrifugal Force  & $f_{V_{\varphi}}^{\hat s} \equiv {{\rho V^2_{\varphi}} \over  {\varpi}}
                   {{{\partial} \varpi} \over {{\partial} s}}$ &  $> 0$ &  $> 0$   \\
                   & $f_{V_{\varphi}}^{\hat n} \equiv {{\rho V^2_{\varphi}} \over  \varpi}
                   {{{\partial} \varpi} \over  {{\partial} n}}$  &  $< 0$   &  $< 0$ \\
\hline
Magn. hoop stress of $B_{\varphi}$ & $T_{B_{\varphi}}^{\hat s} \equiv - {{B^2_{\varphi}} \over {4\pi \varpi}}
                  {{{\partial} \varpi} \over  {{\partial} s}}$  &   $< 0$    &  $< 0$   \\
                  &  $T_{B_{\varphi}}^{\hat n} \equiv - {{B^2_{\varphi}} \over {4\pi \varpi}}
                  {{{\partial} \varpi} \over  {{\partial} n}}$ &   $> 0$    &   $> 0$   \\
\hline
Magn. pressure grad. of $B_{\varphi}$ & $f_{\nabla B_\varphi}^{\hat s} \equiv - {{\partial}
                  \over {{\partial} s}} \left({{B^2_{\varphi}} \over  {8\pi}} \right)$  & $>0$ or $<0$  
& $>0$ or $<0$  \\
                     & $f_{\nabla B_\varphi}^{\hat n} \equiv - {{\partial} \over
                  {{\partial} n}} \left( {{B^2_{\varphi}} \over {8\pi}} \right)$  & $>0^{\rm a}$ & $>0^{\rm a}$ \\
\hline 
Poloidal inertial force & $f_{V_p}^{\hat s} \equiv -\rho V_p
{\partial V_p \over  \partial s} $ & $<0$   & $<0$ \\
                         & $f_{V_p}^{\hat n} \equiv -\frac{\rho V_p^2}{R_c}$ &   $<0$   & $>0$ \\
\hline
Magn. hoop stress of $B_p$ &  ------  &   ------ &  ------ \\
                           & $T_{B_p}^{\hat n } \equiv {{B_p^2} \over {4 \pi R_c}}$  & $>0$   & $<0$ \\
\hline
Magn. pressure grad. of $B_p$ &------  &  ------ &   ------ \\
                             & $f_{\nabla B_p}^{\hat n} \equiv - {{\partial} \over {{\partial} n}}
                              \left( {{B_p^2} \over {8 \pi}} \right)$  & $<0$    & $>0$  \\
\hline
\end{tabular}
\vskip 0.1 true cm
$^{\rm a}$ Only for $B_{\varphi}$ monotonically increasing with $\varpi$
\end{table*}

\noindent$\bullet$
{\it Centrifugal `force'}: ${\vec f}_{V_{\varphi}}= {\hat \varpi}(\rho
V^2_{\varphi}/\varpi)$.  

This centrifugal `force' is always directed outwards, in the
$+{\hat \varpi}$ direction with components :
\begin{eqnarray}
\label{fsvphi}
f_{V_{\varphi}}^{\hat s} \equiv {{\rho V^2_{\varphi}} \over  {\varpi}}
                   {{{\partial} \varpi} \over {{\partial} s}}= \cos \chi (R, \alpha )
{2\lambda^2 \over R G^2 M^2}\left[ {M^2 - G^2 \over 1 - M^2}\right]^2
\nonumber\\
\times \left( 1 -\frac{F}{2}\right)  \alpha
\,,\;\;\;\;\;
\end{eqnarray}
\begin{eqnarray}
\label{fnvphi}
f_{V_{\varphi}}^{\hat n}  \equiv {{\rho V^2_{\varphi}} \over  \varpi}
                   {{{\partial} \varpi} \over  {{\partial} n}}   = \sin \chi (R, \alpha )
{4\lambda^2 \over FRG^2 M^2}\left[{M^2 - G^2 \over 1 - M^2}\right]^2
\nonumber\\
\times  \left[ \frac{R^2}{G^2} +
\alpha \left( \frac{F}{2} - 1\right) \right]
\,.\;\;\;\;\;
\end{eqnarray} 
We see that always $f_{V_{\varphi}}^{\hat s} > 0$, such that the centrifugal `force' acts 
towards accelerating the flow, as a `bead sliding along a rotating wire' in the picture of Blandford \& 
Payne (1982). On the other hand, we have  $f_{V_{\varphi}}^{\hat n}  < 0$ such that the centrifugal `force' 
tends to further decollimate the streamlines, for all angles $\theta$ when the poloidal streamlines 
are deflected from radiality towards the polar axis ($\chi < 0$, or, $0<F<2$) and also for small 
polar angles $\theta$ satisfying $\tan^2\theta < 2/(-F)$, or, $\tan^2\chi < (-F)/2$ 
when the streamlines flare towards the equator ($\chi > 0$, or, $F<0$).

\noindent$\bullet$
{\it Toroidal magnetic force, $\vec f_{B_\varphi} \equiv (\vec \nabla 
\times \vec B_\varphi) \times \vec B_\varphi$}.

The toroidal magnetic force $\vec{f}_{B_\varphi}$ has two components, a tension
$\vec{T}_{B_\varphi}$ and a toroidal magnetic pressure gradient,
$\vec{f}_{\nabla B_\varphi}$, both appearing in Eqs. (\ref{sequation}) and (\ref{nequation}).

\noindent$\bullet$
{\it Magnetic hoop stress of $B_{\varphi}$ }:  
${\vec T}_{B_{\varphi}}= {\hat \varpi}(B^2_{\varphi}/4\pi \varpi)$.

The tension of the toroidal magnetic field is in a direction opposite to 
the centrifugal `force', i.e., it acts in the $- {\hat \varpi}$ direction with components :
\begin{eqnarray}
\label{fsbphi}
T_{B_{\varphi}}^{\hat s} \equiv - {{B^2_{\varphi}} \over {4\pi \varpi}}
                  {{{\partial} \varpi} \over  {{\partial} s}} = - \cos \chi (R, \alpha )
{2\lambda^2 \over R G^2 }\left[ {1 - G^2 \over 1 - M^2}\right]^2
\nonumber\\
\times \left( 1 -\frac{F}{2}\right)  \alpha
\,,\;\;\;\;\;
\end{eqnarray}
\begin{eqnarray}
\label{fsbphi2}
T_{B_\varphi}^{\hat n}  \equiv - {{B^2_{\varphi}} \over {4\pi \varpi}}
                  {{{\partial} \varpi} \over  {{\partial} n}} = - \sin \chi (R, \alpha )
{4\lambda^2 \over FRG^2}\left[{1 - G^2 \over 1 - M^2}\right]^2
\nonumber\\
\times \left[ \frac{R^2}{G^2} + \alpha \left( \frac{F}{2} - 1\right) \right]
\,.\;\;\;\;\;\;\;\;
\end{eqnarray}
Thus, we see that always $T_{B_{\varphi}}^{\hat s} < 0$, such that the hoop stress acts 
towards decelerating the flow. On the other hand, we have  $T_{B_{\varphi}}^{\hat n}  > 0$ such 
that the toroidal magnetic stress tends to collimate the streamlines when the poloidal streamlines 
are deflected from radiality towards the polar 
axis ($\chi < 0$, or, $0<F<2$) and also for small polar angles $\theta$ satisfying 
$\tan^2\theta < 2/(-F)$, or, $\tan^2\chi < (-F)/2$ when they flare towards the equator 
($\chi > 0$, or, $F<0$).

\noindent$\bullet$
{\it Magnetic pressure gradient of $B_{\varphi}$}:
${\vec f}_{\nabla B_{\varphi}}= -\nabla (B^2_{\varphi}/8 \pi)$.

This force is positive or negative depending on 
whether $B_{\varphi}$ decreases or increases with 
${\hat \varpi}$, respectively. Its components are:

\begin{eqnarray}
\label{fsbphi3}
f_{\nabla B_\varphi}^{\hat s} \equiv - {{\partial}
                  \over {{\partial} s}} \left({{B^2_{\varphi}} \over  {8\pi}} \right)  = -\cos \chi (R, \alpha )
{2\lambda^2 \over R G^2 }{1 - G^2 \over (1 - M^2)^2}
\nonumber\\
\times \left[ \left( \frac{F}{2} - 1\right) \left( G^2 + 1 \right) +  R {1 - G^2 \over 1 - M^2}{ {d} M^2 \over {d} R}
 \right] \alpha
\,,\;\;\;\;
\end{eqnarray}
\begin{eqnarray}
\label{fnbphi}
f_{\nabla B_\varphi}^{\hat n} \equiv - {{\partial}
                  \over {{\partial} n}} \left({{B^2_{\varphi}} \over  {8\pi}} \right)
= -\sin \chi (R, \alpha )
{4\lambda^2 \over FR G^2 }{1 - G^2 \over (1 - M^2)^2} 
\nonumber\\
\times \left\{ \frac{R^2}{G^2}(1-G^2) 
+ \left[\frac{FR}{2}{1 - G^2 \over 1 - M^2}{ {d} M^2 \over {d} R}
\right.\right.\nonumber\\
\left.\left. +\left( \frac{F}{2} -1 \right) \left(F-G^2+1\right) \right] \alpha\right\}
\,.\;\;\;\;\;\;\;\;
\end{eqnarray}
In the accelerating initial region  $B_{\varphi}$ decreases with $R$ and thus with $\varpi$
 so the positive gradient of the toroidal magnetic field accelerates the flow.
In the case of collimated flows $B_{\varphi}$ increases with $\varpi$ so
the negative gradient of the toroidal magnetic field acts in the same direction as
the hoop stress, collimating and decelerating the wind. Note that it is easy 
to see that the meridional components of the tension and the
gradient of $B_{\varphi}$ have an equal amplitude.

\noindent$\bullet$
{\it The poloidal flow inertial and curvature forces,} 
$\vec f_{V_p} \equiv - \rho
({\vec V_p} \cdot \nabla) {\vec V_p}$. 

The two components of this force are: 
\begin{eqnarray}
\label{fsvp}
f_{V_p}^{\hat s} \equiv
- \rho V_p { \partial V_p \over  \partial s}
= - {2\cos \chi (R, \alpha ) \over G^4} \left\{  \left[
{ {d} M^2 \over  {d} R} + \frac{M^2}{R}(F-2)\right ]\right. 
\nonumber\\
+ \alpha   \left[ \frac{G^2}{R^2} \left( \frac{F^2}{4} -1 \right)
{ d M^2 \over d R} \right. \nonumber\\
\left. \left.+ {M^2G^2\over 4R^3 }\left(
{F^3\over 2} -2 F^2 -2F+8
+RF{ {d}F\over {d}R}
\right) \right] \right\}
\,.\;\;\;\;\;\;\;\;
\end{eqnarray}
\begin{eqnarray}
\label{fnvp}
f_{V_p}^{\hat n} \equiv
- \frac{\rho V_p^2}{R_c}
= \sin \chi (R, \alpha ) \frac{2M^2}{FRG^4} \left[
\left( R{ dF\over dR} -\frac{F^2}{2} +F\right)
\right. \nonumber\\
\left. + {\alpha G^2\over R^2} \left(
{F^3\over 4} -F - R{ {d}F\over {d}R} \right)\right]
\,,\;\;\;\;\;\;\;\; 
\end{eqnarray}
where $R_c$ is the local radius of curvature. 
Along a poloidal streamline, $f_{V_p}^{\hat s}$ balances  
all other forces along $\hat s$ on the poloidal plane.  
Obviously, when the poloidal flow inertial 
force $-f_{V_p}^{\hat s} $ is positive the flow is accelerated, 
as in the inner region, while when it is negative 
the flow decelerates, as when oscillations are present. 
Perpendicularly to a poloidal streamline, i.e. along $\hat n$, 
the poloidal flow curvature force $-f_{V_p}^{\hat n}$ 
is positive when the streamlines turn towards the axis 
(as in the inner region), zero in the radial case 
and negative when there is some flaring away from 
radiality (e.g., in the oscillating zone).

\noindent$\bullet$
{\it Poloidal magnetic force, $\vec f_{B_p} \equiv (\vec \nabla 
\times \vec B_p) \times \vec B_p$}.

This force is directed only normal to the poloidal magnetic
fieldlines, since its $ \hat s-$component does not appear in 
Eq. (\ref{sequation}). 
As for $B_{\varphi}$, the poloidal magnetic force has two components, a tension
$T^{\hat n}_{B_p}$ and a poloidal magnetic pressure gradient,
$f^{\hat n}_{\nabla B_p}$, both appearing in Eq. (\ref{nequation}).

\noindent$\bullet$
{\it Magnetic hoop stress of $B_p$: $T^{\hat n}_{B_p}$}
$\equiv \hat n \cdot ({\vec B_p} \cdot \nabla) {\vec B_p}/ 4 \pi$.

This force is focusing the streamlines towards the axis when they collimate and
towards the equatorial plane when they are flaring:  
\begin{eqnarray}
\label{fnbp}
T_{B_p}^{\hat n} \equiv
\frac{B_p^2}{4\pi R_c} = - \sin \chi (R, \alpha ) \frac{2}{FRG^4} \left[
\left( R{\hbox{d}F\over\hbox{d}R} -\frac{F^2}{2} +F\right)
\right.\nonumber\\
\left. + {\alpha G^2\over R^2} \left(
{F^3\over 4} -F - R{ {d}F\over {d}R} \right)\right]
\,.\;\;\;\;\;\;\;
\end{eqnarray}
Note that in the regime of the oscillations this
poloidal magnetic curvature force is in opposite phase with the poloidal
velocity curvature force. However, the ratio of their amplitude
equals to $M^2$ and thus in highly superAlfv\'enic outflows 
($M \gg 1$), the poloidal magnetic curvature
force is negligible in comparison to the poloidal flow curvature
force.

\noindent$\bullet$
{\it Magnetic pressure grad. of $B_p$: $f_{\nabla B_p}^{\hat n}$}
$\equiv -\hat n \cdot \vec \nabla \left( {B_p}^2/8\pi\right)$. 

The pressure gradient of the poloidal magnetic field reads as: 
\begin{eqnarray}
\nonumber
f_{\nabla B_p}^{\hat n} \equiv
- { {\partial} \over  {\partial} n}\left( \frac{B^2_p}{8\pi} \right )
= \sin \chi (R, \alpha )\left\{
{2(2-F) \over RG^4} \right.\nonumber\\
\left. - \frac{4}{FG^2R^3}\left(
\frac{F^2}{4} -1\right)
 - {\alpha \over G^2R^2}\left[
 \frac{F}{2} { {d}F\over {d}R}
\right.\right.\nonumber\\
\left.\left.+\frac{2(F-2)}{R}\left(\frac{F^2}{4} -1 \right) \right]
 + \frac{4\alpha}{FR^5}
\left(\frac{F^2}{4} -1 \right)
\right\}
\,.\;\;\;\;
\end{eqnarray}
In an almost radial expansion $B_p$ drops
rather fast with $R$,  $B_p \sim R^{-2}$ and thus 
$f_{\nabla B_p}^{\hat n}$ has a strong component 
along the radial direction. In such cases, the gradient 
of the poloidal magnetic pressure has a
component along -$\hat n$ acting to decollimate the outflow. 
In fact, this strong poloidal magnetic pressure 
gradient when combined with
the rather weak poloidal magnetic curvature force for
superAlfv\'enic flows, results in a total poloidal magnetic force
which has roughly the magnitude and sign of $\vec
f_{\nabla B_p}^{\hat n}$, Figs. \ref{f9}-\ref{f12}.

\end{document}